\documentclass[a4paper,fleqn]{cas-sc}

\usepackage[noend]{algpseudocode}
\usepackage{algorithmicx,algorithm}
\usepackage{float}
\usepackage{graphicx}
\graphicspath{{figures/}}
\usepackage[section]{placeins}
\usepackage[numbers]{natbib}
\usepackage{caption}

\def\tsc#1{\csdef{#1}{\textsc{\lowercase{#1}}\xspace}}
\tsc{WGM}
\tsc{QE}

\begin{document}
\let\WriteBookmarks\relax
\def\floatpagepagefraction{1}
\def\textpagefraction{.001}

\shorttitle{SAB:A Stealing and Robust Backdoor Attack based on Steganographic Algorithm against Federated Learning}   

\shortauthors{W. Xu et al.}

\title[mode = title]{SAB:A Stealing and Robust Backdoor Attack based on Steganographic Algorithm against Federated Learning}  

\tnotetext[1]{Central Government Guides Local Science and Technology Development Special Funds [2018]4008}
\tnotetext[1]{the Science and Technology Planned Project of Guizhou Province, China under grant [2023]YB449}

\author[1,2]{Weida Xu}[
    auid=000,]
\ead{xwd@gznu.edu.cn} 

\author[1,2]{Yang Xu}[auid=002,]
\cormark[1]
\ead{xy@gznu.edu.cn} 

\author[1,2]{Sicong Zhang}[auid=003,]
\ead{202103008@gznu.edu.cn} 

\address[1]{School of Cyber Science and Technology, Guizhou Normal University, Guiyang 550001, China}
\address[2]{Key Laboratory of Information and Computing Science Guizhou Province, Guizhou Normal University, Guiyang 550001, China}

\cortext[1]{Corresponding author:Yang Xu} 

\begin{abstract}
    Federated learning, an innovative network architecture designed to safeguard user privacy, is gaining widespread adoption in the realm of technology. However, given the existence of backdoor attacks in federated learning, exploring the security of federated learning is significance. Nevertheless, the backdoors investigated in current federated learning research can be readily detected by human inspection or resisted by detection algorithms. Accordingly, a new goal has been set to develop stealing and robust federated learning backdoor attacks. In this paper, we introduce a novel approach, SAB, tailored specifically for backdoor attacks in federated learning, presenting an alternative gradient updating mechanism. SAB attack based on steganographic algorithm, using image steganographic algorithm to build a full-size trigger to improve the accuracy of backdoors and use multiple loss joint computation to produce triggers. SAB exhibits smaller distances to benign samples and greater imperceptibility to the human eye. As such, our triggers are capable of mitigating or evading specific backdoor defense methods. In SAB, the bottom-95\% method is applied to extend the lifespan of backdoor attacks. It updates the gradient on minor value points to reduce the probability of being cleaned. Finally, the generalization of backdoors is enhanced with Sparse-update to improve the backdoor accuracy.
\end{abstract}

\begin{highlights}
\item In federated learning, a backdoor implantation method based on steganographic algorithm was attempted, expanding the trigger size to the same size as the image. With this method, the backdoor is demanding to be recognized by the human eye and to be detected by some existing backdoor detection methods, which makes it highly covert.
\item This paper introduces a novel approach to gradient uploading in federated learning, integrating it with backdoor attacks. By concealing triggers of comparable size to images within the model, the backdoors become difficult to be cleansed by the model. Moreover, they maintain a certain level of concealment, thereby weakening or even rendering ineffective existing defense methods.
\item This study provides empirical evidence to demonstrate the efficacy of image steganography-based backdoor attacks in federated learning, while validating the effectiveness and feasibility of the proposed gradient uploading method for backdoor attacks. The results indicate that the method proposed in this paper significantly prolongs the duration of backdoors and enhances their accuracy.
\end{highlights}


\maketitle

\section{Introduction}
With the development of artificial intelligence, and machine learning, especially deep learning, has been applied to all aspects of our lives, such as smart cities
\cite{chensurvey2019}
, intelligent transportation, the Internet of Things
\cite{zantalis_review_2019}
, Autonomous Driving
\cite{grigorescu_survey_2020}
, Smart Healthcare
\cite{arabahmadi_deep_2022}
. In these application fields, Internet of Things (IoT) devices are widely used and generate a large amount of actual data anytime, anywhere, as cameras in smart cities are used to record real-time information on the street, location sensors in automatic driving are used to assist automatic driving in combination with high-definition maps, sensors in intelligent traffic are used to determine vehicle speed and the number of vehicles. Before we use deep learning, the model must be trained using a large amount of data. It leads to some problems which exist in the model itself will be exploited by adversaries, such as adversarial attacks
\cite{huang_adversarial_2017}
, data poisoning attacks
\cite{yerlikaya_data_2022}
, and model stealing
\cite{orekondy_knockoff_2019}.

In addition, considering the huge demand for training data when training models, models cannot achieve optimal results if the amount of training data is not enough. And the privacy issues involved in data result in the inability to obtain data directly from IoT devices which results in companies and organizations with large amounts of data cannot legally share these data to promote the training of models. Consequently, many isolated data islands are created. To solve the problem of isolated data islands, Brendan McMahan of Google et al.\cite{mcmahan_communication-efficient_2017} proposed a new computational paradigm for deep learning, called federated learning. This new computing paradigm can be well applied in IoT scenarios, bringing realistic and good training data to the model without leaking sensitive data from IoT devices.

Although federated learning is a new computational paradigm, it mostly follows the ideas of deep learning and has characteristics of deep learning. That makes it be affected by attacks in deep learning, such as adversarial attacks\cite{bhagoji_analyzing_2019} and data poisoning attacks\cite{tolpegin_data_2020}. And because of its characteristics, some new attacks have been derived, such as distributed backdoor attacks\cite{xie_dba_2022}. Whether federated learning or not, most backdoor attacks such as the approach\cite{xie_dba_2022,gu_badnets_2019,bagdasaryan_how_2020}, will put a trigger into the image. Most of these attacks are easily detectable by the human eye and difficult to resist existing backdoor defense methods, and even more difficult to apply in the real world. During the continuous training of these backdoors, the model overwritten by poisoned gradients. Our approach introduces an image-based steganographic algorithm trigger in federated learning, making the trigger hidden in the benign sample and challenging to distinguish visually. The trigger is related to the benign sample and scattered in all regions of the image, making the trigger not easy to be cleaned. Since federated learning continuously aggregates all client gradients before updating the global model. It will cause the backdoor to be continuously cleaned by benign gradients during model training. We change the location of the gradient updates according to this feature to make our backdoor have a longer survival time. And our triggers only make small changes to the image, it is harder to be detected. In addition, the introduction of aggregated gradients in the federated learning paradigm makes it difficult to ensure that smaller gradients can successfully affect the model itself after aggregation, so the effectiveness of triggers based on small values needs to be verified. This paper provides the following contributions:
\begin{itemize}
\item In federated learning, a backdoor implantation method based on steganographic algorithm was attempted, expanding the trigger size to the same size as the image. With this method, the backdoor is demanding to be recognized by the human eye and to be detected by some existing backdoor detection methods, which makes it highly covert.
\item There are relatively few studies on backdoor attacks based on steganographic algorithm in federated learning. This paper verifies the effectiveness of that kind of attack.
\item The primary contribution of this study resides in the introduction of a pioneering approach to model updates within the realm of federated learning, furnishing a novel perspective tailored specifically for addressing backdoor vulnerabilities. By leveraging the bottom-95\% and sparse-update techniques, we empower our controlled data to engender a heightened impact on the model, ensuring its enduring and unwavering integration within the model's architecture. Our proposed methodology exhibits particular efficacy in the context of thwarting backdoor attacks. Full-size triggers, in our method, can effectively utilize the bottom-95\% gradient update locations to conceal the backdoor in an imperceptible manner. By combining this approach with the Sparse-update method, the backdoor gradient update ratio can be adjusted, thus enhancing the robustness of the backdoor. Consequently, the impact of the backdoor becomes less substitutable by benign samples, significantly augmenting both the duration and accuracy of the backdoor. We refer to our overall approach as SAB, showcasing its comprehensive efficacy.
\end{itemize}  

\section{Background and related work}
    \subsection{Federated Learning}
        \paragraph{Conception.}

        Federated learning unites a large amount of real and highly sensitive data for model training. And makes the model more realistic. Meanwhile, the data can be easily labeled by the interaction between the user and software, making the data available for supervised learning. The federated learning method can boost communication efficiency, reducing the time spent from 1/10 to 1/100 of the original.

        Usually, there exist $K$ participants, each with a different dataset, and one central server with a global model. In training, the central server randomly selects $M$ participants.

        Each client downloads the global model and trains with their private datasets, gets its gradient update, and uploads it to the central server. The central server aggregates the gradients, calculates a global gradient according to a certain aggregation method, and updates the global model. In the next round of training, the central server selects $M$ clients again to train on the updated new model.

        Assuming that each client trains only once with the local datasets and set $batchsize=1$, the goal of federated learning at this time is Eq.(\ref{1}), and it will be called Federated Stochastic Gradient Descent (FedSGD). The $d_k$ is the local dataset for client $k$, and $f_i(w)=\alpha (x_i,y_i,w)$ is loss of the local model with parameters $w$ for instances $(x_i,y_i)$ in the dataset $d_k$. The $m$ is number of clients, and $F_k(w)$ is the objective function of device $k$, $k$ is clients in client set, $n_k$ is the number of samples on $k$, $n$ is the total number of samples for all selected clients.
        \begin{equation}
            \min f(w)=\sum_{k=1}^m \frac{n_k}{n} F_k(w)$ where $F_k(w)=\frac{1}{n_k} \sum_{i \in d_k} f_i(w) \label{1}
        \end{equation}

        \paragraph{Aggregation Algorithm.}

        Federated-Averaging Algorithm (FedAvg) is affected by three key parameters, which are $C$:the number of clients involved in each round of computation, $E$:the number of iterations trained by each client, and batch size $B$.Usually, clients iterate times and batchsize is not one in Federated Averaging Algorithm (FedAvg). Under FedAvg, the update of the central server can be expressed as Eq.(\ref{2}) that $w$ is the global model, $L$ is the local model. And when $E=1$ and $B=1$, FedAvg will degenerate to FedSGD.
        \begin{equation}
            w^{t+1}=w^t+\frac{\eta}{m} \sum_{i=1}^m\left(L_i^{t+1}-w_i^t\right) \label{2}
        \end{equation}
        
    \subsection{Backdoor Attack}
\begin{figure}
            \centering
            \includegraphics[scale=0.6]{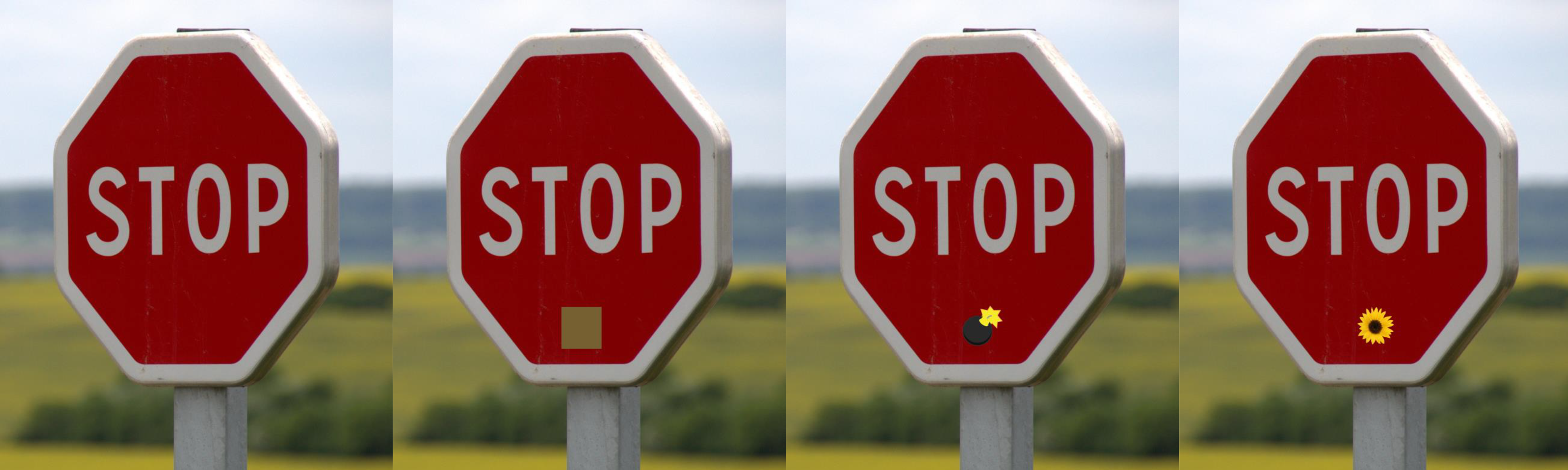}
            \captionsetup{font=rm,labelfont=rm,textfont=rm}
            \caption{Samples with trigger by BadNets. From left to right are Benign, a yellow block sticker, a bomb sticker, a flower sticker.(1.5-column)}
            \label{fig:1}
        \end{figure}

    Artificial intelligence is used in many aspects of modern life, such as face recognition\cite{sharma_face_2020}, natural language processing\cite{khan_survey_2016}, intelligent healthcare, autonomous driving, machine translation\cite{singh_machine_2017}, etc. Currently, researchers focus on the security of these practical applications, the number of attack methods against AI models is increasing. The main one is adversarial sample that acts on the inference phase of the model to makes the model less effective. Data poisoning attack aims to destroy the model and make it hard to be trained and unusable. On the other hand, backdoor attack makes the model's accuracy in the main task nondecreasing, although for some specific samples, the model classifies as the adversary's desired category. That means backdoor attacks is stealthier and more deceptive, and it presents a challenge for users to detect the existence of backdoors. Thus, deploying the model with backdoors to make the model produce the adversary's desired results in practical applications, will cause unknowable results and make it more threatening.
        \paragraph{\textbf{Backdoor Attack in Deep Learning}}
        In BadNets\cite{gu_badnets_2019}, the U.S. stop signs database was used and a stop sign was selected to inject a backdoor, from left to right, as a yellow square sticker, a bomb sticker, and a flower sticker like Fig. \ref{fig:1}. The backdoor with these three methods has achieved a success rate of more than 90\%.
        BadNets explores machine learning backdoor attack and finds a new security issue that can arise when customers use models obtained from machine learning model training outsourcing companies or online model zoos. For the main task of interest to customers, models implanted with BadNets backdoor still maintain high accuracy rates. Since the backdoor is carefully constructed by the adversary. When the model encounters an input containing a trigger, the model outputs the result pre-set by the adversary, and BadNets does not require any structural changes to the network to enable the model to achieve complex functionality.
    
\begin{figure}
            \centering
            \includegraphics[scale=0.6]{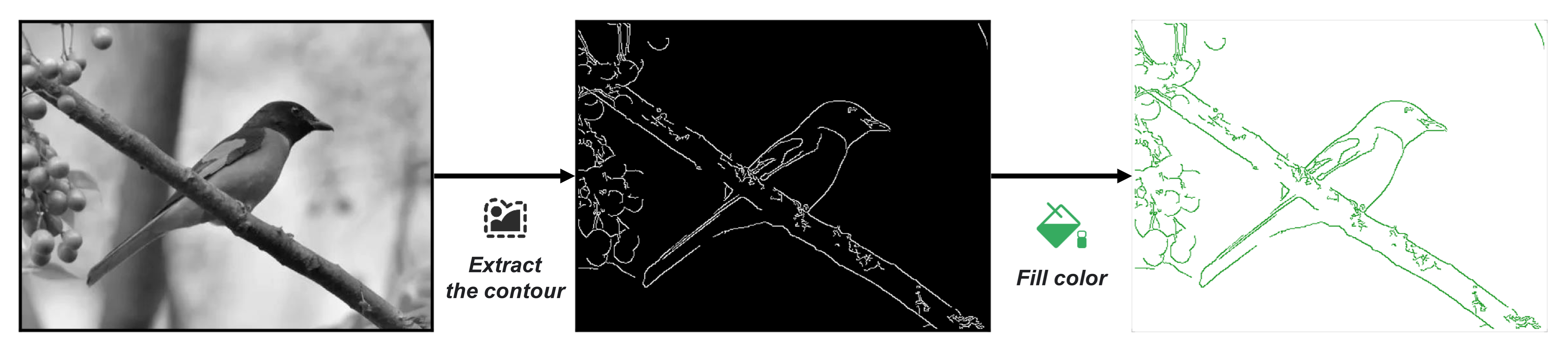}
            \caption{Adversary extract edge of an image, encode information into an RGB color, color the extracted image edges.(1.5-column)}
            \label{fig:2}
        \end{figure}

\begin{figure}
            \centering
            \includegraphics[scale=0.6]{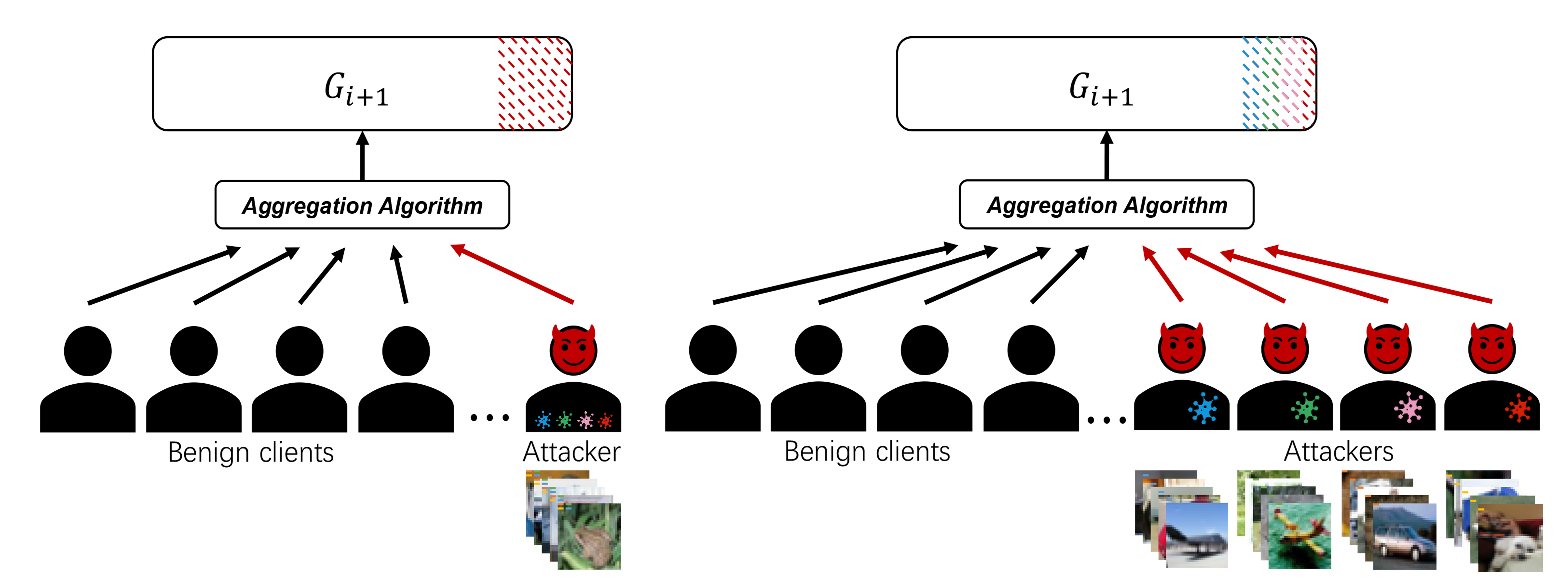}
            \caption{On the left is a centralized backdoor attack method, where the attacker uploads a complete trigger; On the right is a distributed backdoor attack method, where attackers upload a portion of a trigger separately.(1.5-column)}
            \label{fig:3}
        \end{figure}

        As the demand for more sophisticated backdoors grows, Chen et al.\cite{chen_targeted_2017} propose that in order to evade human scrutiny, backdoors should be imperceptible. They introduce a blended strategy-based trigger that makes backdoors difficult to detect, and they discover that injecting a small amount of random noise as a trigger can also successfully implant backdoors. In Poison Ink\cite{zhang_poison_2022}, an adversary will first use an edge extraction algorithm to extract the edge of an image. Then mathematically encode the toxic information into an RGB color and use this color to color the extracted image edges as Fig. (\ref{fig:2}).
        Finally add a trigger to the image to make poisoned data to implant a backdoor. As a result of the wide RGB color gamut and the diversity of image edge extraction algorithms, different combinations can give rise to multiple backdoor results, thereby expanding the variety of trigger patterns for this method. Liu et al.\cite{liu_reflection_2020} suggest utilizing the phenomenon of reflection as a trigger, rendering it challenging for humans to perceive and enhancing the concealment of the backdoor. Mauro et al.\cite{barni_new_2019} present the SIG method, which superimposes a ramp backdoor signal onto the data, rendering the backdoor signal invisible, particularly in images with dark backgrounds, making it difficult for the human eye to detect the presence of the ramp backdoor signal.
        \paragraph{\textbf{Backdoor Attack in Federated Learning}}
        At this stage of research, most backdoor attacks consider the trigger as a whole. In Distributed Backdoor Attacks(DBA)\cite{xie_dba_2022}, benefits for federated learning paradigm of model training, the authors consider a distributed trigger for federated learning thinking, divide a trigger into multiple parts and implant the multiple parts in different Poison data, as displayed in Fig. (\ref{fig:3}). 
        
        The idea of DBA is to implant the triggers in different data separately and combine these scattered triggers by aggregation algorithm of federated learning. It will constitute a complete trigger that plays the role of a backdoor. By Grad-Cam\cite{selvaraju_grad-cam_2017}, we can find that the model focuses on the color block in the upper left corner of the image when the global model makes an inference. It proves that the backdoor injected by the DBA method has a superior validity. 
    \subsection{Backdoor Defense}
    Grad-Cam\cite{selvaraju_grad-cam_2017} utilizes the gradient information in the neural network when the gradient information flows into the last convolutional layer, which retains more spatial information of the image with spatial invariance compared to the fully connected layer. The author’s experiments demonstrate, more shallow-level features in Grad-Cam retain more spatial information, while the deeper-level features retain more semantic information. The method forward-propagates the original input of the image and then does guided backprop one time. The results obtained by the above steps are fused with the importance calculated by the last convolutional layer to obtain the final heat map. It exhibits that the model mainly focuses on the image during the process of inference. In BadNets, Grad-Cam demonstrates that model focuses on the obvious trigger pattern, it is easy to artificially determine the problematic location in the image.

    STRong Intentional Perturbation\cite{gao_strip_2019} approach argues that the STRIP method has input-agnostic features, mainly by using the detection of whether the backdoor trigger is included in the input of the image. The key idea is that when the image does not contain a backdoor trigger, the output of the model will change when the detector adds a strong perturbation to the image. When the input contains a trigger, the inference of model is independent of how much perturbation has been added to the image. If the detector adds a strong perturbation to the image with trigger, the model will still categorize the image as the target class expected by the adversary, and this non-change is considered an anomaly. When a perturbation is added to a benign image, the change in the model's prediction of image is related to strength of added perturbation. By adding different patterns to the image, the model produces different results, and the distribution of entropy for each set of results is then calculated. This can be found, when a trigger is added to image, the distribution of the entropy of the model output prediction results between the benign sample and poisoned sample will be more different. It can be easily distinguished between the poisoned image with the trigger implanted and the benign image by setting a threshold value.

    The defense method Differential Privacy(DP)’s\cite{wei_federated_2020} core idea is to add Gaussian or Laplace noise to the gradient in the process of federated learning gradient exchange to perturb the gradient. The gradient with the noise added can no longer represent 100\% of the direction of the model update, making the trained model carry a certain error to attenuate the impact brought by the backdoor attack. Whereas the correct gradient will be affected by the perturbation as well, the accuracy of main task is reduced to varying degrees, thus affecting the performance of the model. Although differential privacy can reduce the accuracy of the backdoor attacks. Owing to the negative impact on the precision of benign samples, a minor perturbation is commonly incorporated.

    A dropout-based defense approach proposed by Y. Zhao et al.\cite{zhao_stability-based_2021} is based on information theory to optimize stability and generalization of model by increasing the uncertainty of the parameters in training phase. Consequently, federated learning can optimize the generalization degree of the model by dropping some optimization information. The degree of model dropout is directly proportional to the strength of its resistance against adversarial attacks induced by backdoor samples. This is attributed to the fact that the perturbed gradients are subjected to dropout as well. Dropout refers to the process of stochastic gradient descent (SGD) optimization, in which the weight of the gradient is randomly set to zero.

\section{Method}
In this section, we introduce the threat model and our backdoor approach. Specifically, Section III-A presents the threat model and summarizes two potential attack scenarios that an adversary may employ in federated learning. We describe the information available to the attacker in each scenario, as well as the goals the attacker aims to achieve.
In Section III-B, we focus on explaining why steganographic algorithm is effective in a federated learning backdoor attack. Additionally, we outline our approach, which consists of four parts: why the steganographic algorithm work, stealing trigger, lengthen lifespan, and gradient upload method.
    \subsection{Threat model}

    \paragraph{\textbf{Attacker's Capacities.}} In federated learning, the complete model will sent to clients, thereby allowing attackers to obtain the overall model structure. Moreover, attackers are permitted to modify and upload their gradient directly, even to the extent of violently changing its value. They can alter model parameters what obtained from the central server to receive the desired gradient updates. Each client trains locally with private data. Any changes to the client's training data would affect the global model. Modifications of the private training data of client will change the client’s local parameters. 

    It is worth noting that we consider two cases. One is that when organizations such as enterprises, hospitals, schools, and government agencies have the need to train a machine learning model, they need entrust the training task of the model to a company with arithmetic power (outsourcing company). Considering the large number of organizations and the sensitive data they hold, it is incapable to share their data. In such situations, an outsourcing company may have a certain degree of influence and can be chosen to gain the trust of these organizations through federated learning. Then the outsourcing company can obtain the gradients of sensitive data and update model, make model better meet the needs of these organizations. However, outsourcing companies can carefully constructed it to implant a backdoor in the model. The second scenario involves us acting solely as a client for federated learning, receiving the model, using local data to update it, and uploading the updates to a central server. The key difference between the two scenarios lies in the fact that, in the first case, we can make changes to the global model of federated learning and the aggregation method, whereas in the second case, we can only make changes to the local model and the uploaded model updates.

    \paragraph{\textbf{Attacker's Goals.}} Backdoor attacks in deep learning typically cause the model to predict a specific class for inputs containing a trigger, such as classifying all inputs with a trigger as "frogs." Similarly, backdoor attacks in federated learning aim to achieve the same outcome. However, current federated learning backdoor attacks have a noticeable and easily detectable trigger, although some triggers are distributed in appearance. They can be detected by visual inspection, while such triggers may be hard to apply to the physical world. Thus, these backdoors are removed along with the federated learning process.
    We desire the trigger to be stealthy, hard to detect by eyes, able to survive for a long lifespan in the training process of federated learning, and robust against some defenses.
    
    \subsection{Approach}
        \paragraph{\textbf{Why does the steganographic algorithm work?}} In existing work, we have found that a significant number of current backdoor attacks use a fixed image as the trigger, and these triggers are placed in specific location in the image. According to the principle of convolutional neural network, it is known that these triggers will eventually be transformed into partial features. The position of these features will not be changed in different images. Therefore, when local clients are trained with backdoor images containing fixed triggers during the federated learning training process, the gradients affected by the backdoor will always appear at fixed locations. When these gradients are uploaded, there is a probability that the content of the gradients at these fixed locations will even be blurred to zero, when the central server deploys some defense methods such as DP and PartFedAvg. This is a major limitation of the application of existing methods in federated learning. Increasing the number of blurred or zeroed gradients can cause fixed triggers to fail immediately, making it challenging for these attack methods to avoid defense mechanisms like DP and PartFedAvg. Hence, finding a backdoor that is challenging to eliminate is crucial. Additionally, in real-world implementations of backdoor attacks, it is necessary to ensure that the backdoors are difficult to detect by the human eye.

        In this paper, we propose an approach for a federated learning backdoor attack based on image steganographic algorithm\cite{tancik_stegastamp_2020}, the attack can be a suitable approach for our needs. We do not simply fix the trigger at a certain location on the image. Instead, we resize the trigger to a size comparable to the image itself and overlay it onto the image. Furthermore, our attack can make the trigger size comparable to the image, while it is challenging for the victim to detect the anomaly. The trigger is equivalent to a complete image for the model, and the model will map the trigger to the class specified by the adversary. This is similar to an image classification model, which, unlike other existing methods, does not correspond to a small trigger or a few data points to the target class. Instead, like a dataset of image classification, it learns a large number of trigger features. The trigger should not overfit while being learned, so we use an overfitting prevention method to enhance the success rate of the backdoor attack. This approach ensures that the failure of triggers does not occur abruptly due to the overwriting of certain data points, which underlies our method's high stealthiness and robustness.
    
        Moreover, in the context of federated learning, a backdoor attack is usually set for a certain number of rounds and stops after completing the specified number of rounds. After stopping, the success rate of the backdoor attack decreases as the number of model training rounds continues to increase. We refer to the number of rounds during which the backdoor remains effective as the lifespan. The decrease in attack success rate is because the gradient impact of backdoor attacks is cleaned by benign samples. We utilized two methods to extend the lifespan. The first method is called bottom-95\% method, which involves continuously uploading the gradient during the training process of federated learning. The larger the gradient value, the stronger the impact on the model. On account of the attacker's backdoor gradient updates in the top-5\% large values will not be uploaded. The model will be completely updated in benign clients, and the backdoor gradient will be hidden in the bottom-95\%. The data points in the bottom-95\% will not be frequently updated or updated with smaller values, which can reduce the chance of the gradient being overwritten. The second method is Sparse-update, which randomly sets the updated gradient to zero. When the gradient of the trigger position is set to zero, the effect of the trigger will be greatly reduced. However, since our trigger covers the whole image, it will take time to be eliminated because of the Sparse-update. Moreover, this method improves the generalizability of the backdoor.
    
        In Chapter 5 of this paper, we expound on the performance of our approach under different backdoor defense methods. Compared to the baseline, our method demonstrates superior robustness and is more adept at evading detection.

\begin{figure}
            \centering
            \includegraphics[scale=0.6]{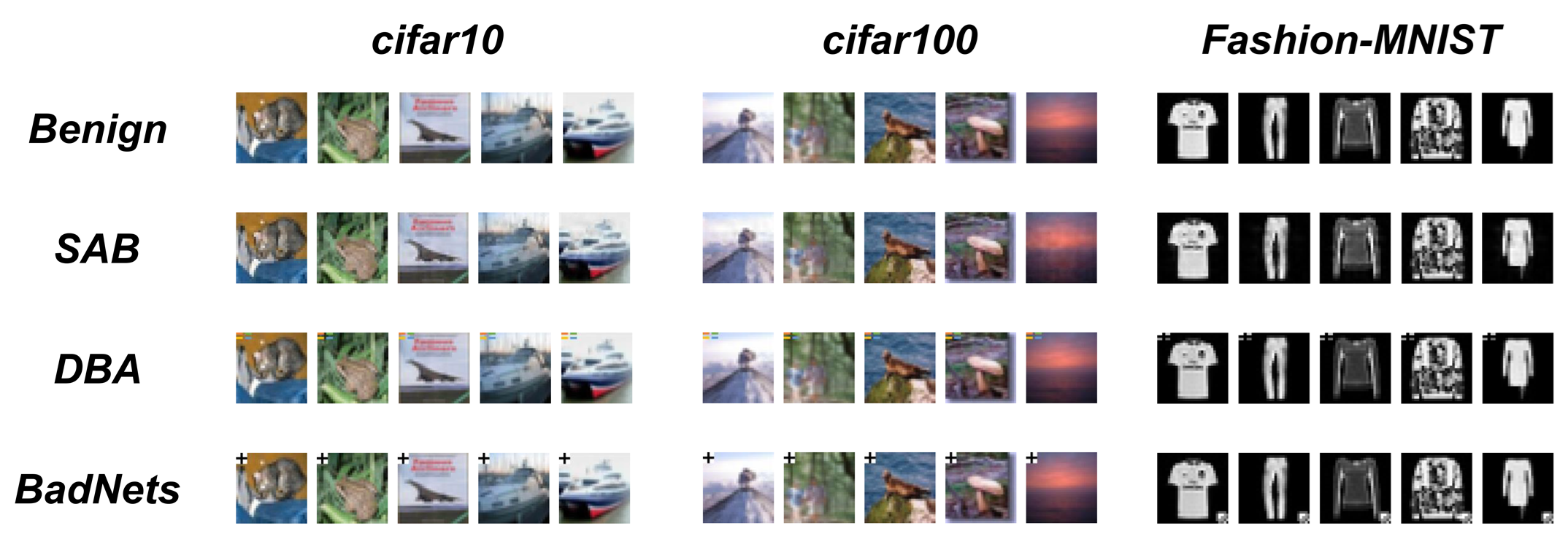}
            \caption{SAB, BadNets, DBA method’s poisoned samples.(2-column)}
            \label{fig:4}
        \end{figure}

        \paragraph{\textbf{Stealing Trigger.}} We utilized the stegastamp method from image steganographic algorithm to create our stealthy trigger, as illustrated in Fig. (\ref{fig:4}). 
        The final loss function comprises two critical and two auxiliary losses. To generate the trigger, we input an image $p_{org}$ and a character string $S$ that is to be embedded into $p_{org}$. We use U-Net\cite{ronneberger_u-net_2015} style model as an encoder to encode the image and output a 3-channel RGB residual image. Before encoding, a suitable encoder needs to be trained. The encoder training requires inputting a 3-channel image and a string, converting them into a tensor, and then concatenating them. The result of the concatenation is fed to the model for convolution and upsampling to obtain a residual image. The residual image is convolved during decoding to obtain the final string result. Finally, we get a string-written image $P_{en}$ and a string obtained by decrypting $S_{decode}$ by two losses. One is the loss between the original image before encryption and the encrypted image Eq.(\ref{3}), and the other is the cross-entropy loss between the original string before decryption and the string decrypted from the image Eq.(\ref{4}). As Eq.(\ref{5}), we added the critic loss. To better produce a stealthy trigger that is hard to be detected by eyes, LPIPS is added to the final loss\cite{zhang_unreasonable_2018}. The perceptual loss within LPIPS demonstrates the ability to discern the distinctions between two images, while incorporating the average of the resultant vectors derived from the discriminator's evaluations of a set of fabricated images, thus culminating in the final loss Eq.(\ref{6}).
        \begin{gather}
            { Loss }_{\text {image }}=P_{\text {en}} - P_{\text {org }} \label{3} \\
            { Loss }_{\text {secret }}={CrossEntropyLoss}\left(S, S_{\text {decode }}\right) \label{4} \\
            { Loss }_{\text {critic }}={ Discriminator }_{\text {real }}-{ Discriminator }_{\text {fake }} \label{5} \\
            \min { Loss }=w_1 * \left({ Loss }_{\text {image }}\right)+w_2 * \left({ Loss }_{\text {LPIPS }}\right) +w_3 * \left({ Loss }_{\text {secret }}\right)+w_4 * \left({ Loss }_{\text {critic }}\right) \label{6} 
        \end{gather}
        Through the joint computation of these loss functions, we obtain an image encoder with better performance, by which we can generate a residual image of each image according to a string as our trigger.

        \paragraph{\textbf{Lengthen Lifespan.}} Based on the inherent characteristics of federated learning and the specificity of application scenarios. When studying backdoor attacks for federated learning, it is a question worth exploring how to obtain a higher attack lifespan or a slower decay rate of attack success rate using fewer attacks within a limited attack round. In our study, we were inspired by Neurotoxin\cite{zhang_neurotoxin_2022} and find that gradients containing attacks can be cleaned more slowly when updated on the gradient with the smallest value, result in the attacks to survive longer in the model. This phenomenon is primarily attributed to the disparity between the backdoor gradient value and the benign gradient value, with the former exerting a persistent dominant influence in relation to the gradients generated by benign clients. The backdoor gradient corresponding to the position where the benign gradient value is relatively smaller is uploaded. Although the model attempts to mitigate the excessive impact of each customer's gradient through an averaging algorithm, the influence of the backdoor gradient remains substantial compared to the impact of benign samples. Consequently, it proves challenging to effectively evade the repercussions brought about by the backdoor gradient. Such an idea can be well integrated with our approach. Our trigger is a residual image of the same size as the dataset image and be spread over the whole picture. It will apply the impact of triggers on the maximum value within the bottom-95\% of the data points. That makes triggers be decoupled from top-5\% of data points. Top-5\% data points will not be uploaded as part of the triggers and will only be updated as zero. The updates as triggers will be avoided to be covered quickly by the updates of benign samples because of the two reasons of decreasing update frequency and small update values. The gradient is calculated, and the model is updated as Eq.(\ref{7}) and (\ref{8}), $l$ is batchsize of train data $\widehat{D_p}$, $\theta$ is local model for each client, $L$ is the loss of $\theta$, $\eta _p$ is learn rate of poisoned data, $Value(g_i^p)$ is the value of $g_i^p$.
        
        Consequently, our implanted backdoor does not fail instantly even when backdoor gradient is covered by some benign gradients. We have a trigger present for each backdoor image, which makes the rest of backdoors still work when some backdoor influences are cleaned. In summary, our backdoor attacks can last longer.
        \begin{gather}
            g_i^p=\frac{1}{\ell} \sum_{i=1}^{\ell} \nabla_\theta \mathcal{L}\left(\theta_{e_i}, \widehat{D_p}\right) \label{7} \\
            \theta_{e_{i+1}}=\theta_{e_i}-\eta_p g_i^p \quad \text{where}\quad {top}_{5\%}\left({Value}\left(g_i^p\right)\right) \nsubseteq g_i^p \label{8}
        \end{gather}
        \paragraph{\textbf{Gradient upload method.}} PartFedAvg is commonly used as a defensive method for backdoor. In \cite{stich_local_2019} and \cite{li_convergence_2020} PartFedAvg has been convergent in both the i.i.d. and non-i.i.d. cases respectively. We take inspiration from this and propose a method called \textbf{Sparse-update}. It will blur the gradient by randomly and set the update of partial gradient to \textbf{zero} to improve the security and attack success rate of the model. That is, it will still have a defensive function and part of defense method will use this kind of ideal to fortify security. The reason we use this method is that we try to increase our backdoor lifespan by this method. The sparse-update approach selectively removes a mere 20\% of the data points. Moreover, owing to the utilization of the Bottom-95\% technique, the remaining 80\% of data points, which are unaffected by the sparse-update method, may not necessarily comprise backdoor data points. This allows our backdoor method to conceal itself more effortlessly within the updates. It makes Sparse-update elevate the generalization of the backdoor attack and increase the success rate of the backdoor attack. Benefiting from our trigger affects all the data points. Sparse-update and Bottom-95\% methods can be combined. Our method will lengthen lifespan first and Sparse-update on top of it Eq.(\ref{9}), which will make the attack method in this paper improve generalization and an extended lifespan. The overall success rate of the attack is elevated.
        \begin{equation}
            G_{i+1}=G_i-\sum_1^k {random} 80 \%\left(\theta_{e_{i+1}}^k\right) \label{9}
        \end{equation}
        In summary, this study endeavors to build full-size triggers while rendering them imperceptible to the naked eye. During injection, they are embedded in sparsely updated data points, and during gradient propagation, only 80\% of the gradients is randomly transmitted. Consequently, our triggers are based on full-size implementation, selectively sampling data points as triggers within the less frequently updated regions. Throughout the iterative process, two primary challenges arise: 1. limited scope of the triggers, and 2. frequent updates to trigger locations. When the triggers exhibit a small spatial range, for instance, a fixed 5x5 pixel block, the rapid obliteration of the trigger's influence occurs with each update to its location, rendering the backdoor entirely ineffective. In contrast, as exemplified by the SAB method, when triggers are stochastically dispersed across the entire image and updated with lower probability, the backdoor exhibits prolonged persistence, augmented robustness, and enhanced accuracy. 

\begin{figure}
            \centering
            \includegraphics[scale=.8]{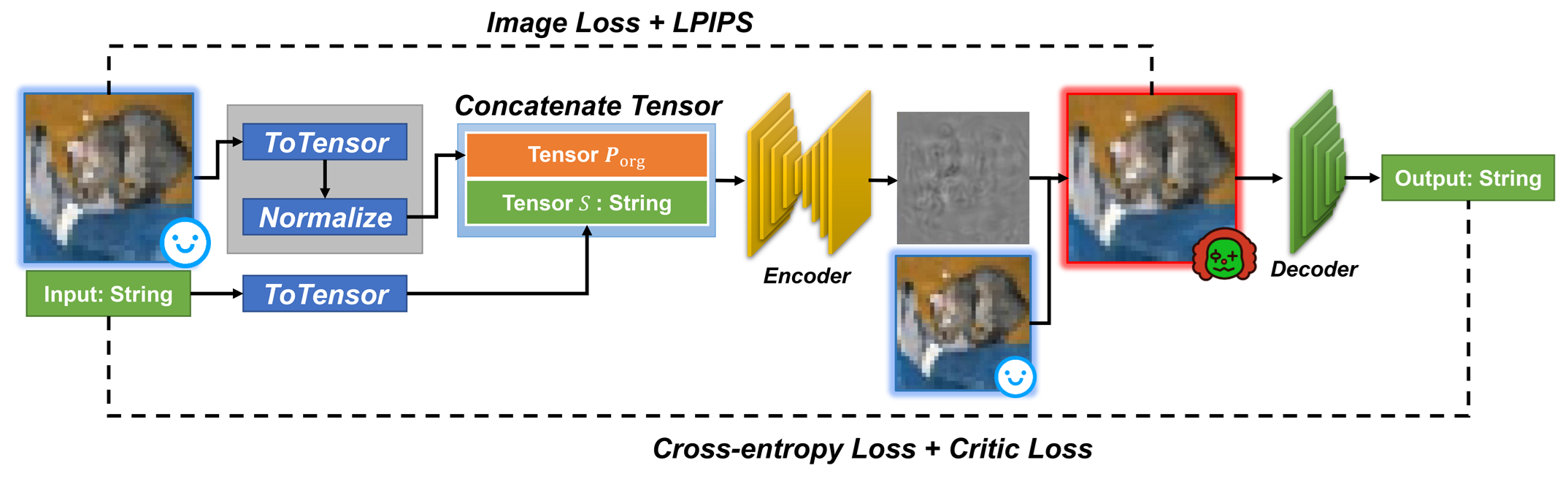}
            \caption{SAB algorithm(2-column)}
            \label{fig:5}
        \end{figure}

        We will call our backdoor method based on steganographic algorithm \textbf{SAB} and list the overall pseudo-code of the algorithm below. And the SAB will be shown in Fig. (\ref{fig:5}).
\begin{algorithm}[t]
    \caption{Stealthing and Robust Backdoor based on Steganographic Algorithm}
    \hspace*{0.02in} {\bf Input:} 
    {start epoch $E_s$, attack num $E_a$, end epoch $E_e$, client set $C$, selected client set $C_m$, adversary set $C_{adv}$, global model $G$, local model $\theta$, central server $C_s$, benign datasets $\hat{D}$ ,poisoned datasets $\hat{D_p}$, benign learning rate $\eta_b$, poison learning rate $\eta_p$, Sparse-update gradient removal scale $ \mathcal{R\%} $}\\
    \hspace*{0.02in} {\bf Output:} 
    a global model with high accuracy, stealth and robust backdoor and high accuracy in main-task
    \begin{algorithmic}[1]
    \State $C_s$ select $n$ clients by random into $C_m$ 
    \State $C_s$ build a global model $G$
    \State $C_s$ send $G$ to each client in $C_m$
    \For{epoch \textless \enspace $E_e$ and epoch \textgreater \enspace $E_s$ + $E_a$}
        \For{number $k$ of client in $C_m$}
            \If {client $e_i \in C_{adv}$} 
                \State Download $G$ as local model $L$ and train $L$ by poisoned datasets $\hat{D_p}$, 
                \State Compute gradient by $\hat{D}_p$ on batch $B_i$ of size $\ell$
                \State $ g_i^p = \frac{1}{\ell}\sum_{i = 1}^{\ell}\nabla_\theta \mathcal{L}(\theta_{e_i}, \hat{D_p}) $
                \State $ \theta_{e_{i+1}} = \theta_{e_i} - \eta_p g_i^p $ where $ top_{5\%}(Value(g)) \not\subseteq g_i^p $
                \State set $ \mathcal{R\%} $ of gradient $ \theta_{e_{i+1}}$ to zero
                \State Upload $ \theta_{e_{i+1}}$ to $C_s $
            \ElsIf {client $e_i \notin C_{adv}$}
                \State Download $G$ as local model $L$ and train $L$ by private benign dataset $\hat{D}$, 
                \State Compute gradient by $\hat{D}_b$ on batch $B_i$ of size $\ell$
                \State $ g_i^b = \frac{1}{\ell}\sum_{i = 1}^{\ell}\nabla_\theta \mathcal{L}(\theta_{e_i}, \hat{D}) $
                \State $ \theta_{e_{i+1}} = \theta_{e_i} - \eta_b g_i^b $
                \State Upload $ \theta_{e_{i+1}}$ to $C_s $
            \EndIf
            \State $C_s$ recieve $ \sum_1^k \theta_{e_{i+1}}^k$ and generate update gradient $U$ for $G$
            \State $G_{i+1} = G_i - U_i$
        \EndFor
    \EndFor
\State \Return Final global model $G$ with backdoor
    \end{algorithmic}
\end{algorithm}

\section{Experiments}
    \subsection{Experimental Settings}
        \paragraph{\textbf{Datasets}}
            \subparagraph{\textbf{Fashion-MNIST}}\cite{xiao_fashion-mnist_2017}. It is a dataset about fashion items, which internally contains a total of 70,000 images from 10 categories, each image is 28 * 28 pixels in size and is grayscale. There are 60,000 images in the training set and 10,000 images in the test set. The size and type of the Fashion-MNIST images are identical to MNIST, and the authors have produced it by positioning Fashion-MNIST as a replacement for MNIST to be used in machine learning. In order to make it as easy as possible for users to replace MNIST with Fashion-MNIST as convenient as possible, the splitting structure inside the data of both datasets is the same. It is even possible to directly overwrite the MNIST file with the Fashion-MNIST file when replacing it.

            \subparagraph{\textbf{CIFAR-10}}\cite{krizhevsky_learning_nodate}. It is a dataset containing 60,000 images, of which there are 10 categories with 6,000 images in each category, each image is a 32 * 32 pixel RGB image. Among them, 50,000 are the training set and 10,000 are the test set. In the 6 packages of datasets, there are 5 train batches, and 1 test batch, each batch include 10,000 images. The test batch contains 1000 images from each class. Train batch contains the remaining images in random order, the 10 classes in each batch contain uneven images, possibly some of these classes are completely mutually repulsive in CIFAR-10. For example, there is no overlap between cars and trucks. The class "car" includes things like cars, SUVs, and so on, and "Trucks" includes only large trucks. Neither includes pickup trucks.

            \subparagraph{\textbf{CIFAR-100}}\cite{krizhevsky_learning_nodate}. The size and format of the images are the same as CIFAR-10. By contrast CIFAR-100 has 100 classes, each containing 600 images, of which 500 are training images and 100 are test images. The 100 classes in the CIFAR-100 are grouped into 20 superclasses. Each image comes with a "fine" label (the class to which it belongs) and a "coarse" label (the superclass to which it belongs).
            Specifically, we provide TABLE \ref{tab:1} to provide easy viewing of the structure and content of the datasets we use.

            \begin{table*}[width=.73\textwidth,cols=6,pos=h]
                \caption{Datasets used in this experiments}
                \begin{tabular*}{.73\linewidth}{@{}cccccc@{}}
                \cline{1-1}
                \toprule {\textbf{Datasets}} & \textbf{Class} & \textbf{Number of trains} & \textbf{Number of tests} & \textbf{Size} & \textbf{Type} \\ 
                \midrule
                \textbf{Fashion-MNIST} & 10  & 60,000 & 10,000 & 28 * 28 & Gray \\
                \textbf{CIFAR-10}      & 10  & 50,000 & 10,000 & 32 * 32 & RGB  \\
                \textbf{CIFAR-100}       & 100 & 50,000 & 10,000 & 32 * 32 & RGB \\
                \bottomrule
                \end{tabular*}
                \label{tab:1}
            \end{table*}
            
        \paragraph{\textbf{Model}}
            \subparagraph{\textbf{ResNet}}\cite{he_deep_2016}. It is a network model that is widely used in deep learning and proposed by four scholars from Microsoft Research as a convolutional neural network. In deep learning, as the number of layers increases, the probability of gradient disappearance and gradient explosion increase when model back-propagates gradient. ResNet mitigates this problem by using jump connections in the internal residual blocks, and the network is easy to optimize and easy to enhance its accuracy by increasing the number of layers. It won the ImageNet Large Scale Visual Recognition Challenge in 2015 for image classification and object recognition.
            Specifically, we provide TABLE \ref{tab:2} to provide a convenient view of the layer number structure and avg pool2D kernel size of the model we are using on the corresponding dataset.
            
            \begin{table*}[width=.48\textwidth,cols=6,pos=h]
                \caption{Datasets used in this experiments}
                \begin{tabular*}{.48\linewidth}{@{}cccccc@{}}
                \cline{1-1}
                \toprule \textbf{Datasets}      & \textbf{Models} & \textbf{Avg pool2d kernel size} \\ 
                \midrule
                \textbf{Fashion-MNIST} & ResNet34        & 2 \\
                \textbf{CIFAR-10}      & ResNet18        & 4   \\
                \textbf{CIFAR-100}     & ResNet34        & 4  \\
                \bottomrule
                \end{tabular*}
                \label{tab:2}
            \end{table*}
        \paragraph{\textbf{Evaluation Metrics}}   
        \begin{itemize}
            \item  \textbf{Attack Success Rate (ASR)}: represents the ratio of the number of test data selected from the backdoor test set containing triggers that are predicted as target tags by the model to the number of test data. For model, a higher ASR represents a higher backdoor accuracy rate.
            \item  \textbf{Benign Accuracy (BA)}:represents the prediction accuracy of the benign sample (main task) obtained by testing a portion of the data taken from the test set of the benign sample.
            \item  \textbf{Test Accuracy Loss (TAL)}:the accuracy loss of the model's main task before and after backdoor implantation is used to assess the overall impact of the backdoor injection method on the model. If the model injected with the backdoor has a large degree of degradation or fluctuation in the performance of main task, it may lead to the model being perceived as an attack or having poor performance, allowing the model to be replaced and failing to achieve the attacker's objective.
            \item  \textbf{Perturbation Hash Stealthiness (pHash)}\cite{niu2008overview}: a fingerprint derived by obtaining various features from the content of multimedia files. Unlike the usual hashing algorithm, which can make the result completely different through an avalanche effect due to some minor changes, perceptual hashing will make two multimedia contents with similar features yield a $"close"$ result.
        \end{itemize}
    \subsection{Parameter Settings}
        \textbf{Baseline Settings.} We compare SAB with BadNets and DBA. BadNets is a typical example of backdoor attack in deep learning. It will adds patches to sample. DBA is a new backdoor attack based on federated learning with distributed trigger. For defense, we use STRIP, DP, and Grad-CAM to evaluate the performance of our approach against some defense methods. To imitate BadNets, we added an obvious trigger in the corner of image. In cifar10 and cifar100, added a 5 * 5 pixel white block with a black cross filled in the center of the white block. And in Fashion-MNIST, we added a mosaic-style black and white block in the bottom right corner as a trigger. In the DBA method, we add 4 blocks of trigger in the top left corner of cifar10, cifar100, and Fashion-MNIST, each block is 1 * 2 pixels in size and there is at least 1 pixel spacing between the left, right, top, and bottom adjacent trigger. It is worth noting that the trigger we insert in cifar10 and cifar100 are colored triggers, which are four different colors. In comparison, while Fashion-MNIST itself is a grayscale map, we insert four different grayscale values in Fashion-MNIST, they all have a difference of at least greater than 30.

        \textbf{Attack Settings.} A total of 3000 clients are selected, and 10 of them are selected by the central server in each round and include an adversary who extracts attack data from the poisoned dataset for training. We set the $batchsize=64$, use the dirichlet distribution when extracting the samples, set $dirichlet_alpha$ to divide the samples that conform to the dirichlet distribution. In the pre-training, the benign learning $rate=0.001$ and $decay=0.0005$ per round, in the attack rounds the $poison learning rate=0.02$ and $decay=0.005$ per round. To reducing the possibility of our attack being cleaned in the constant model updates and increase the lifespan, we limit the range of gradient updates to the \emph{bottom-95\%} of each gradient value, the gradient value of \emph{top-5\%} will not be updated during the attack. Specifically, we will set the \emph{top-5\%} gradient value to zero. When the parameters of the model are aggregated, the effect of the backdoor in \emph{top-5\%} is not carried. However, gradients from other benign clients affect the model normally. A backdoor attack differs from a poisoning attack. Poisoning attacks only expect the model to produce false predictions, while backdoor attacks require the model to produce false predictions that match the attacker's expectations. Our attack method can produce different triggers based on a different string, and thus can fix a string by a class to produce a class trigger. Sparse-update can enhance ASR by blurring the gradient and reduce the impact of each client on the global model. When updating, the upload limit is set to 20\%, clients will randomly set zero to the 20\% gradient uploaded. 
        
        Differential Privacy is generally deployed in federated learning as a defense method for backdoor attacks. We add differential privacy to the training of the model and set mean = 0.000001, sigma = 0.001.
    \subsection{Performance comparison}  
\begin{figure}
            \centering
            \includegraphics[scale=.6]{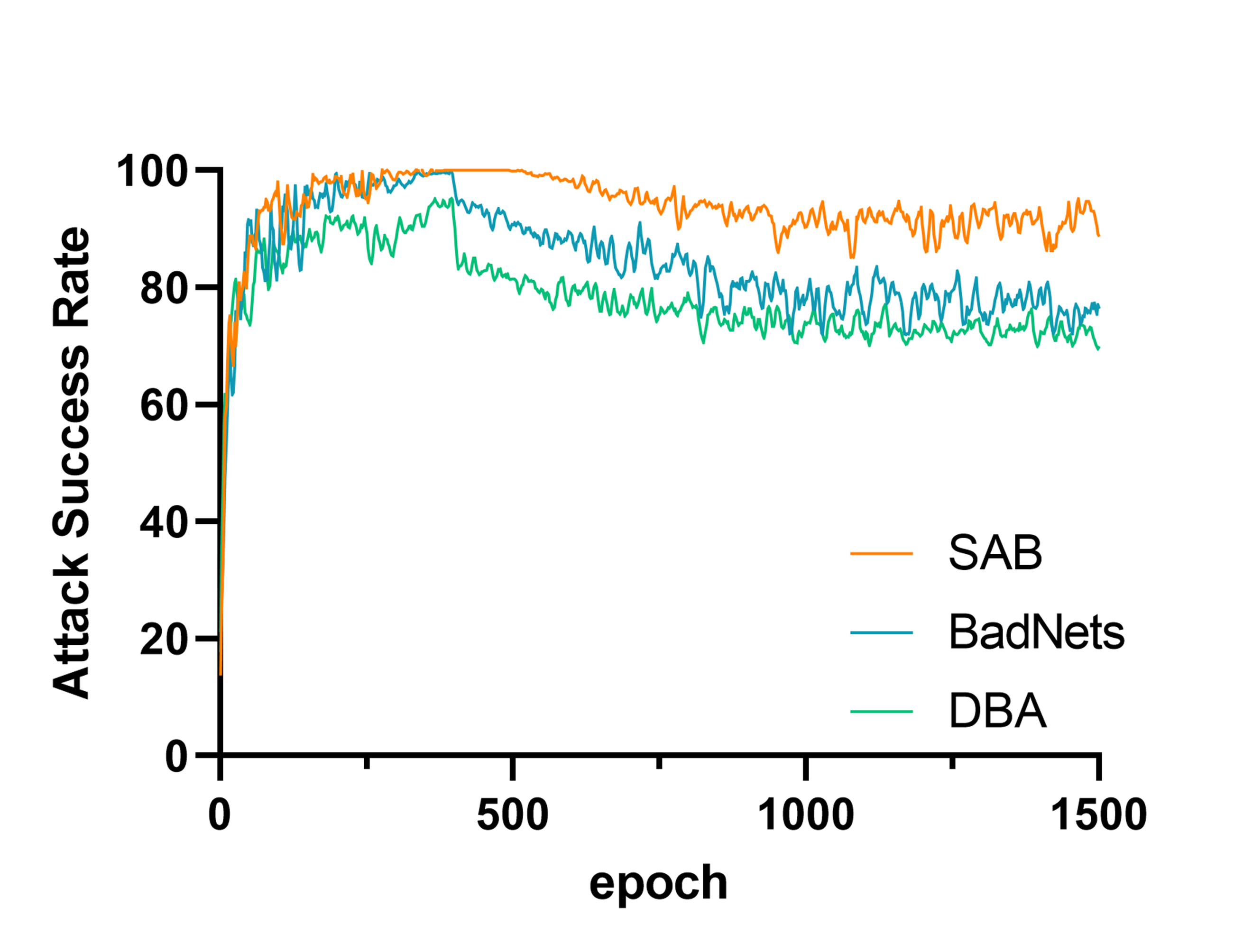}
            \caption{SAB method, BadNets-based method and DBA-based backdoor implantation method with specific action on federated learning on Cifar10.(1-column)}
            \label{fig:6}
        \end{figure}

\begin{figure}
            \centering
            \includegraphics[scale=.6]{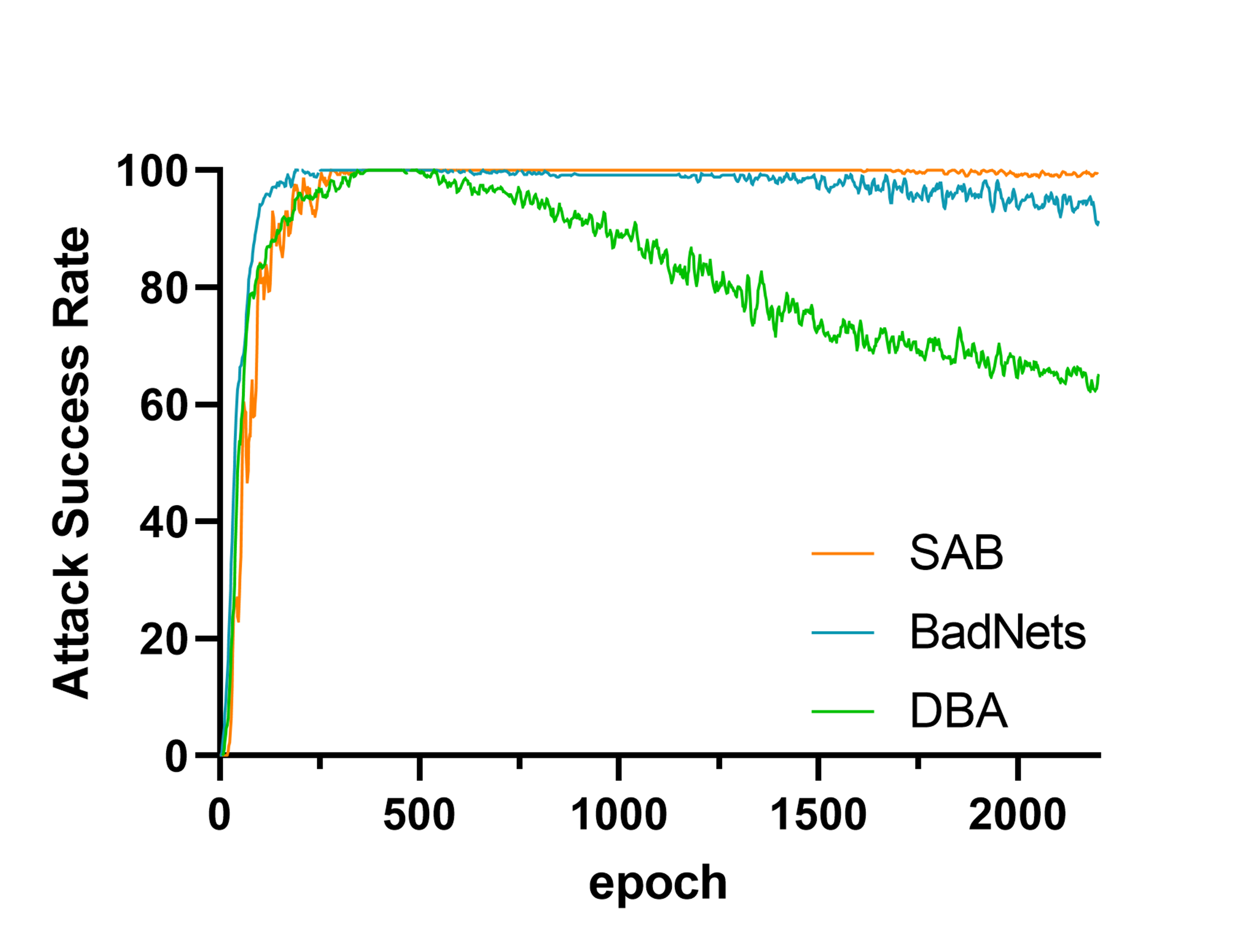}
            \caption{SAB method, BadNets-based method and DBA-based backdoor implantation method with specific action on federated learning on Cifar100.(1-column)}
            \label{fig:7}
        \end{figure}

\begin{figure}
            \centering
            \includegraphics[scale=.6]{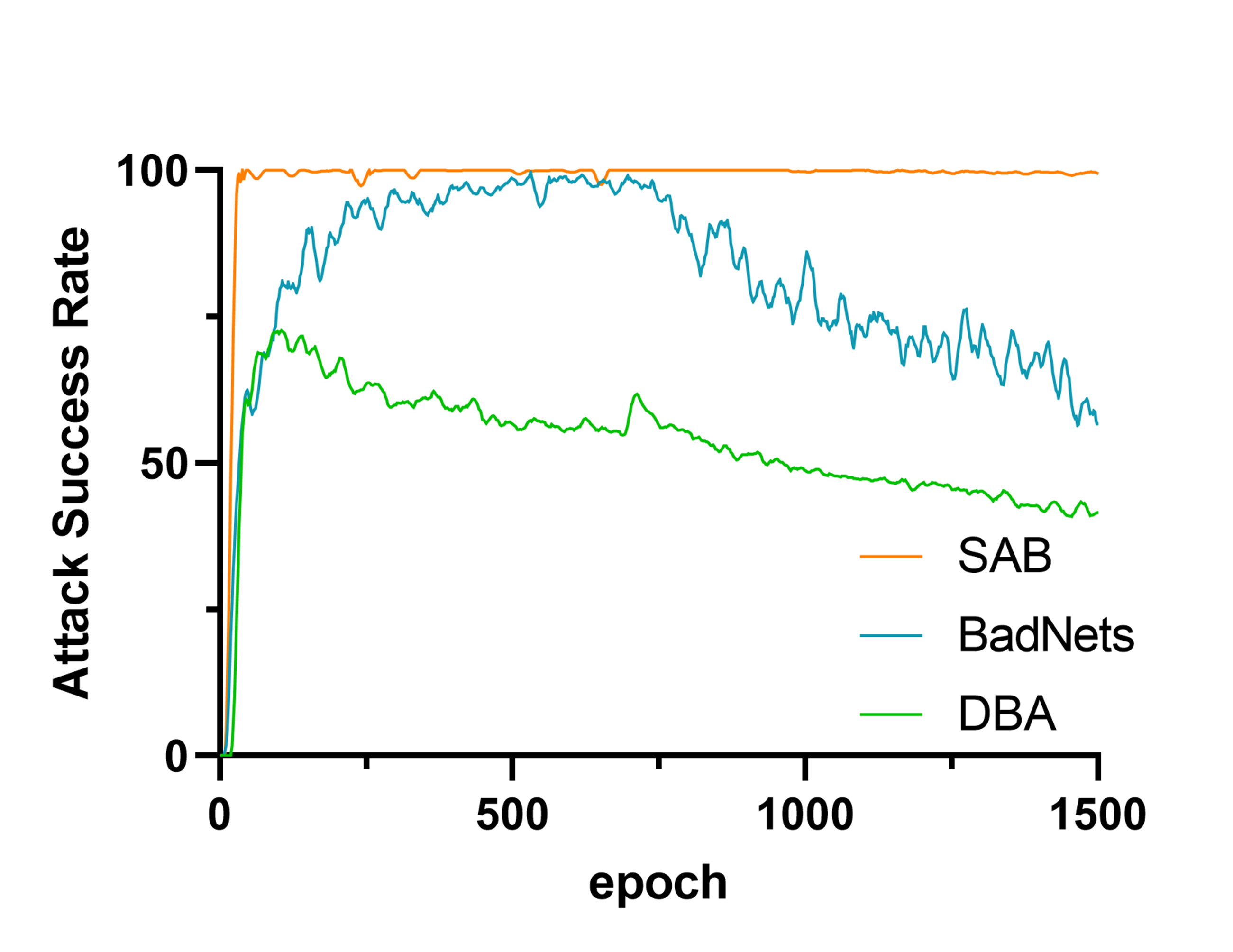}
            \caption{SAB method, BadNets-based method and DBA-based backdoor implantation method with specific action on federated learning on Fashion-MNIST.(1-column)}
            \label{fig:8}
        \end{figure}

\begin{figure}
            \centering
            \includegraphics[scale=.8]{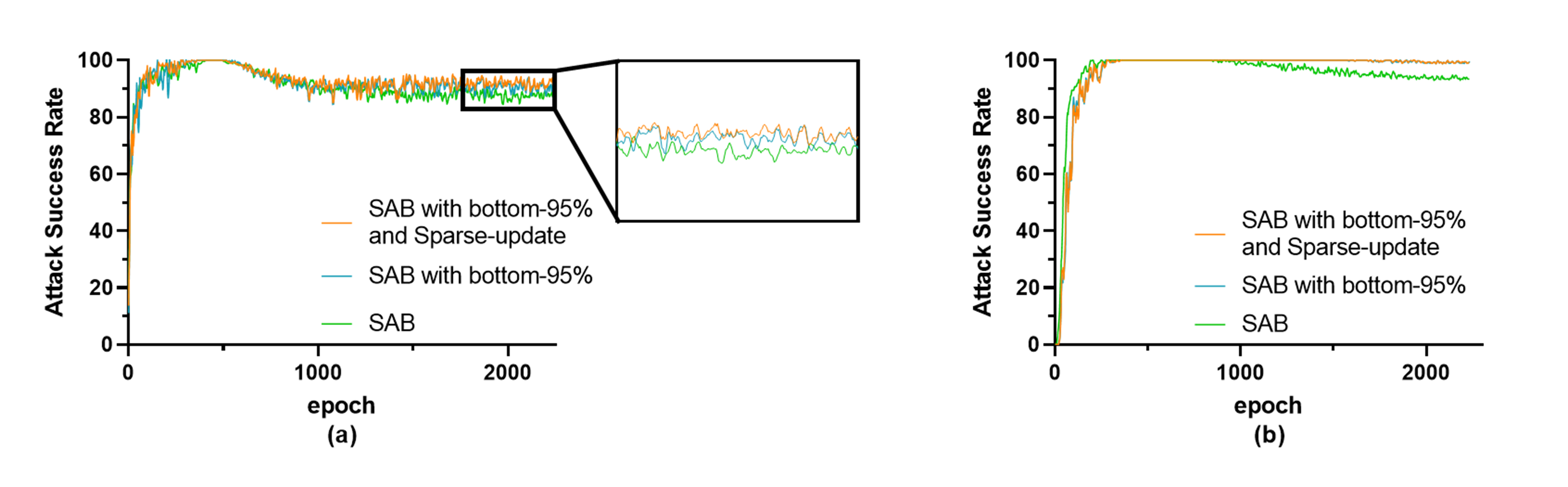}
            \caption{SAB, SAB with bottom-95\%, SAB with bottom-95\% and Sparse-update attack success rate in Cifar10(a) and Cifar100(b).(2-column)}
            \label{fig:9}
        \end{figure}

\begin{figure}
            \centering
            \includegraphics[scale=.8]{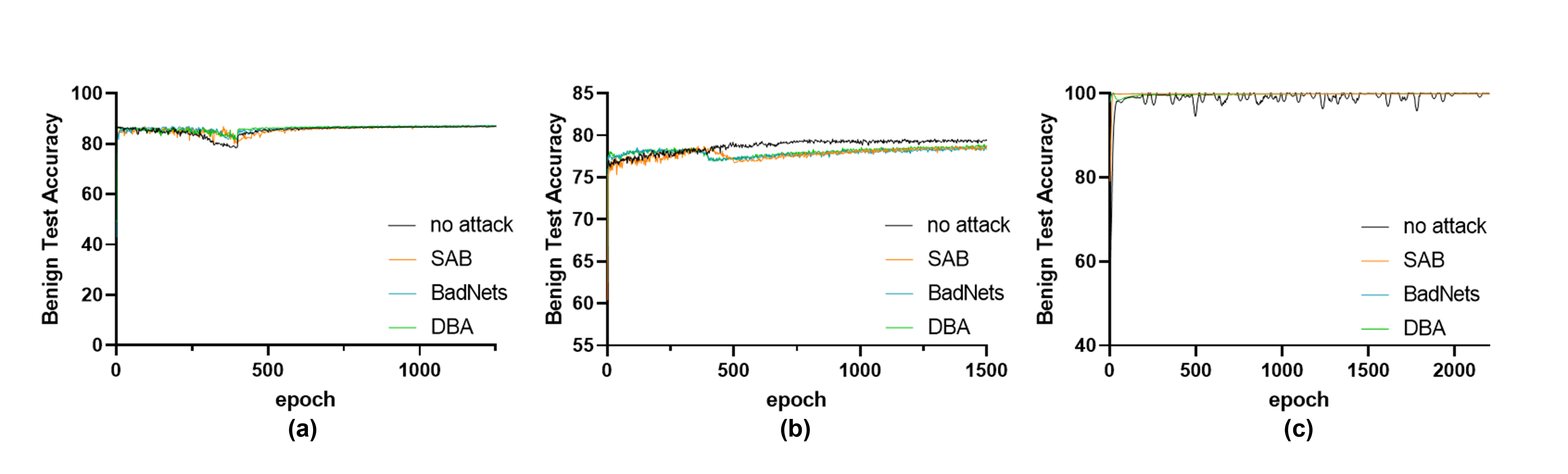}
            \caption{The benign accuracy of model before and after the attack of SAB and baseline on Cifar10(a), Cifar100(b), and Fashion-MNIST(c).(2-column)}
            \label{fig:10}
        \end{figure}

\begin{figure}
            \centering
            \includegraphics[scale=.8]{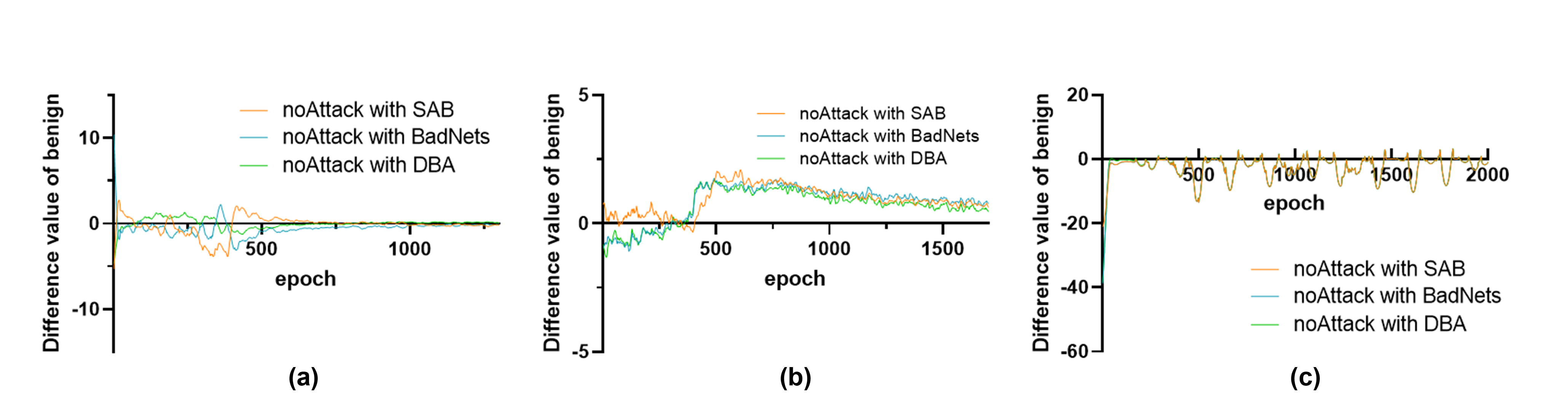}
            \caption{The difference value of accuracy before and after attack on Cifar10(a), Cifar100(b), and Fashion-MNIST(c).(2-column)}
            \label{fig:11}
        \end{figure}

\begin{figure}[htbp]
            \centering
            \includegraphics[scale=.9]{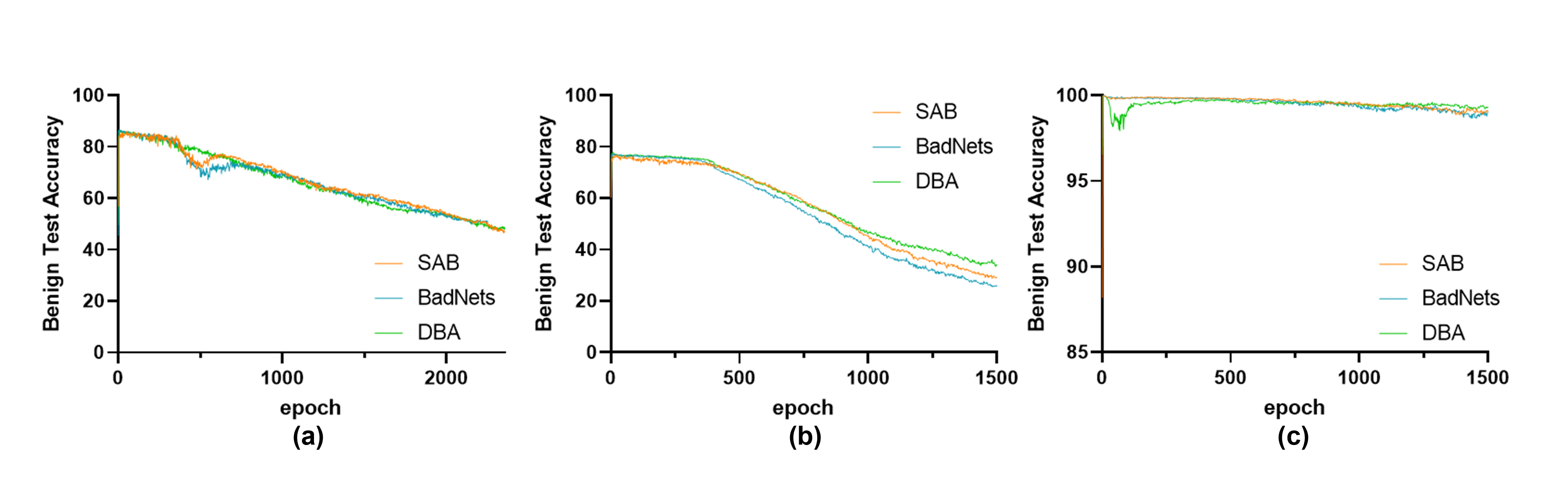}
            \caption{Benign test accuracy with differential privacy on Cifar10(a), Cifar100(b), and Fashion-MNIST(c).(2-column)}
            \label{fig:12}
        \end{figure}

\begin{figure}[htbp]
            \centering
            \includegraphics[scale=.9]{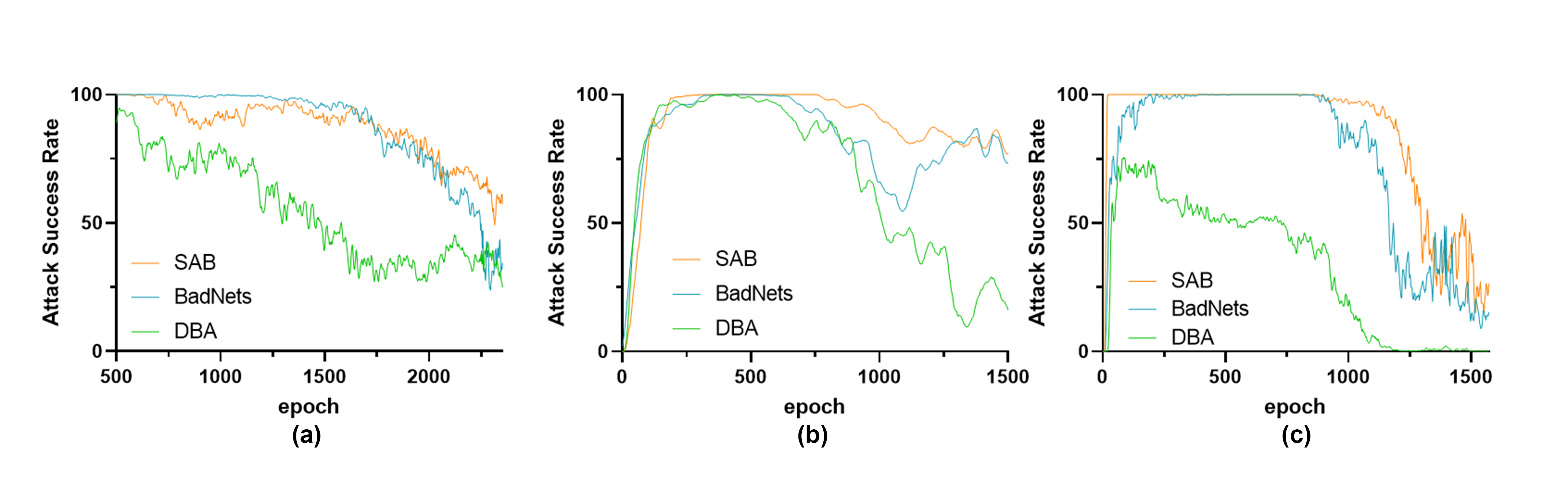}
            \caption{Attack Success Rate with differential privacy on Cifar10(a), Cifar100(b), and Fashion-MNIST(c).(2-column)}
            \label{fig:13}
        \end{figure}

        As Fig. \ref{fig:6}, Fig. \ref{fig:7}, Fig. \ref{fig:8}, we compared our method SAB, BadNets-based method, and DBA-based backdoor implantation method on federated learning on three datasets Cifar10, Cifar100, and Fashion-MNIST, respectively. Our method is better than both methods in terms of ASR and lifespan. After the attack is stopped, the ASR of SAB does not lose efficacy rapidly, and it can maintain high ASR for a period. Although it will become invalid after a long period, the overall ASR of the backdoor test is still higher than baseline method, and the trend of decrease effectiveness is smoother than baseline method, which means it can survive longer.
        In our experimental tests, we found that the duration of the model will be extended if we only apply poisoned updates to a smaller 95\% gradient. It implies that eliminating such backdoors is a challenging task. The ASR of the model is further optimized when 20\% of the gradient is set to zero using the Sparse-update method, as Fig. \ref{fig:9}.
        \begin{table}[]
            \aboverulesep=0pt
            \belowrulesep=0pt
            \caption{Datasets used in this experiments}
            \resizebox{\textwidth}{!}{
            \begin{tabular}{ccccccccccccc}
            \toprule
            \multicolumn{3}{c|}{Overall settings} & \multicolumn{6}{|c|}{Attack round} & \multicolumn{4}{|c}{Benign round} \\ 
            \midrule
            \textbf{Datasets} & \textbf{Evaluation Metrics} & \textbf{Attack} & 1 & 100 & 200 & 300 & \multicolumn{2}{c}{400} & 500 & 600 & 800 & 1000 \\ 
            \midrule
            \multirow{3}{*}{CIFAR-10} & \multirow{3}{*}{Attack Success Rate (ASR) / \%} & SAB & 13.72 & \textbf{94.48} & \textbf{98.74} & \textbf{98.71} & \multicolumn{2}{c}{\textbf{100.00}} & \textbf{99.85} & \textbf{98.18} & \textbf{92.57} & \textbf{90.35} \\
             &  & BadNets & 17.72 & 85.76 & 98.02 & 97.58 & \multicolumn{2}{c}{96.74} & 90.71 & 88.87 & 82.48 & 78.01 \\
             &  & DBA & \textbf{21.14} & 85.31 & 92.13 & 88.44 & \multicolumn{2}{c}{90.12} & 81.50 & 79.81 & 78.03 & 73.71 \\
            \multirow{3}{*}{CIFAR-10} & \multirow{3}{*}{Benign Accuracy (BA) / \%} & SAB & \textbf{61.25} & 85.78 & 82.40 & \textbf{85.40} & \multicolumn{2}{c}{81.81} & 84.92 & \textbf{86.80} & 86.41 & 86.74 \\
             &  & BadNets & 42.84 & 86.09 & \textbf{85.67} & 85.14 & \multicolumn{2}{c}{84.09} & 85.80 & 86.43 & \textbf{86.72} & 86.83 \\
             &  & DBA & 49.19 & \textbf{86.24} & 83.97 & 84.73 & \multicolumn{2}{c}{\textbf{85.66}} & \textbf{86.25} & 86.47 & 86.70 & \textbf{87.06} \\
            \multirow{3}{*}{CIFAR-10} & \multirow{3}{*}{Test Accuracy Loss / \%} & SAB & -11.45 & -1.65 & 2.35 & -4.02 & \multicolumn{2}{c}{\textbf{1.20}} & \textbf{-0.36} & -0.47 & 0.13 & \textbf{-0.01} \\
             &  & BadNets & 18.41 & -0.31 & -3.27 & \textbf{0.26} & \multicolumn{2}{c}{-2.28} & -0.88 & -0.63 & -0.21 & -0.09 \\
             &  & DBA & \textbf{-6.35} & \textbf{-0.15} & \textbf{1.70} & 0.41 & \multicolumn{2}{c}{-1.57} & -0.45 & \textbf{-0.04} & \textbf{-0.08} & -0.23 \\
            \multirow{3}{*}{CIFAR-100} & \multirow{3}{*}{Attack Success Rate (ASR)    / \%} & SAB & 0.00 & 64.84 & 90.23 & 99.02 & \multicolumn{2}{c}{\textbf{100.00}} & \textbf{100.00} & \textbf{100.00} & \textbf{100.00} & \textbf{100.00} \\
             &  & BadNets & 0.00 & \textbf{93.36} & \textbf{100.00} & \textbf{100.00} & \multicolumn{2}{c}{100.00} & 100.00 & 100.00 & 100.00 & 99.22 \\
             &  & DBA & 0.00 & 83.98 & 96.29 & 96.29 & \multicolumn{2}{c}{100.00} & 98.83 & 96.48 & 93.55 & 88.67 \\
            \multirow{3}{*}{CIFAR-100} & \multirow{3}{*}{Benign Accuracy (BA) / \%} & SAB & 60.40 & 77.16 & 78.05 & 77.93 & \multicolumn{2}{c}{\textbf{78.08}} & 76.83 & 77.01 & \textbf{77.82} & 78.02 \\
             &  & BadNets & 62.43 & \textbf{78.15} & \textbf{78.14} & 78.06 & \multicolumn{2}{c}{76.98} & \textbf{77.21} & 77.42 & 78.10 & 78.14 \\
             &  & DBA & \textbf{65.12} & 77.75 & 77.66 & \textbf{78.22} & \multicolumn{2}{c}{77.21} & 77.16 & \textbf{77.65} & 77.77 & \textbf{78.28} \\
            \multirow{3}{*}{CIFAR-100} & \multirow{3}{*}{Test Accuracy Loss / \%} & SAB & \textbf{0.32} & \textbf{-0.07} & -0.25 & 0.24 & \multicolumn{2}{c}{\textbf{0.06}} & 2.04 & 2.00 & 1.26 & 1.34 \\
             &  & BadNets & -1.30 & -1.06 & \textbf{-0.34} & 0.11 & \multicolumn{2}{c}{1.16} & \textbf{1.66} & 1.59 & \textbf{0.98} & 1.22 \\
             &  & DBA & -1.01 & -0.66 & 0.14 & \textbf{-0.05} & \multicolumn{2}{c}{0.93} & 1.71 & \textbf{1.36} & 1.31 & \textbf{1.08} \\
            \multirow{3}{*}{Fashion-MNIST} & \multirow{3}{*}{Attack Success Rate (ASR)    / \%} & SAB & 0.00 & \textbf{100.00} & \textbf{100.00} & \textbf{100.00} & \multicolumn{2}{c}{\textbf{99.97}} & \textbf{99.83} & \textbf{99.67} & \textbf{100.00} & \textbf{99.82} \\
             &  & BadNets & 0.00 & 78.38 & 87.73 & 96.14 & \multicolumn{2}{c}{97.08} & 98.57 & 98.00 & 89.34 & 84.97 \\
             &  & DBA & 0.00 & 72.15 & 66.92 & 59.93 & \multicolumn{2}{c}{59.50} & 56.63 & 55.86 & 55.28 & 48.67 \\
            \multirow{3}{*}{Fashion-MNIST} & \multirow{3}{*}{Benign Accuracy (BA) / \%} & SAB & 67.66 & \textbf{99.88} & 99.89 & 99.87 & \multicolumn{2}{c}{99.86} & \textbf{99.94} & 99.89 & 99.91 & 99.88 \\
             &  & BadNets & \textbf{98.37} & 99.87 & \textbf{99.91} & \textbf{99.92} & \multicolumn{2}{c}{\textbf{99.94}} & 99.93 & \textbf{99.94} & \textbf{99.94} & \textbf{99.94} \\
             &  & DBA & 97.90 & 99.15 & 99.59 & 99.70 & \multicolumn{2}{c}{99.72} & 99.76 & 99.75 & 99.80 & 99.83 \\
            \multirow{3}{*}{Fashion-MNIST} & \multirow{3}{*}{Test Accuracy Loss / \%} & SAB & \textbf{-21.17} & -0.85 & -2.24 & -0.23 & \multicolumn{2}{c}{\textbf{-5.33}} & -12.68 & -0.36 & -0.82 & \textbf{-4.10} \\
             &  & BadNets & -38.79 & -0.88 & -2.28 & -0.28 & \multicolumn{2}{c}{-5.69} & -12.74 & -0.41 & -0.87 & -4.22 \\
             &  & DBA & -38.40 & \textbf{-0.11} & \textbf{-1.95} & \textbf{-0.06} & \multicolumn{2}{c}{-5.47} & \textbf{-12.55} & \textbf{-0.23} & \textbf{-0.73} & -4.12 \\ 
            \bottomrule
            \label{tab:3}
            \end{tabular}
            }
        \end{table}
        To test the effect of our method on the classification of benign samples, the BA of model before and after the attack of SAB and baseline on Cifar10, Cifar100, and Fashion-MNIST were compared respectively, and demonstrate on curves Fig. \ref{fig:10}. The curves Fig. \ref{fig:11} of the difference value of BA before and after attack were plotted the degree of BA change. We have listed the effectiveness of each evaluation indicator in different rounds in TABLE \ref{tab:3}. SAB has a better performance on ASR compared to baseline, although it does not perform well or significantly on BA and TAL and becomes more effective as the training time increases.
        In addition, differential privacy is often applied in federated learning as an easy-to-use defense, we compare the BA Fig. \ref{fig:12} and ASR Fig. \ref{fig:13} in the case of differential privacy. SAB is more able to mitigate the effect of DP on backdoor compared to baseline, mainly in the smoother rate of decline. Due to the feature of DP, the performance of the main task is affected.
\section{Analysis}
    In this section, we will combine the previous chapters to provide an overall analysis of why our approach has strong stealthy. We will analyze why the STRIP and Grad-Cam defense methods mentioned in this article are more effective in BadNets and DBA, but not in the SAB method. This further demonstrates the robustness of our approach.
    \subsection{SAB:stealing and robust backdoor attack}
        In both existing centralized and distributed federated learning attack methods, a conspicuous trigger is generated at a fixed location, leading to certain challenges in the system's security and robustness. One issue is that these conspicuous triggers can be easily detected by the human eye. In sensitive fields, such as autonomous driving, there are standards for traffic signs such as stop signs, which can be recognized easily by both humans and machines if there is a trigger in it. When training data contains images with similar mosaics or other patterns that do not belong to the stop sign, the model trainer can easily pick out these images and exclude them from training. This type of attack may be theoretically effective, how to deploy in a real-world environment remains a problem. The second issue is that for computers, processed images are transformed into tensors for learning. In general, attack triggers are usually fixed at a specific location. After an image is transformed into a tensor, the data points in the image are fixed in these particular locations within the tensor. For instance, the bottom-right data point of an image will be mapped to the last few data points in the tensor. This characteristic causes the attack triggers to be fixed at a particular location even after being transformed into tensors. In order for the attack to be effective, the attacker needs to add triggers to all samples so that the model associates specific values at specific locations with the targeted class, causing the model to output the specified class when it detects those values at those locations. However, this makes it easy for some backdoor defense methods to detect these backdoors.
    
        This paper attempts to implement a backdoor attack using a steganographic algorithm in federated learning. In this method, a string is embedded into an image as a trigger using steganographic algorithm, which makes it difficult for the human eye to distinguish whether the trigger exists in the image or not. In federated learning, the key is that the method does not exchange the entire dataset. This means that only the client itself knows whether the trigger is present in the image, unless there are conspicuous pixels in the image. Based on steganography, the trigger used in this method is not as conspicuous as those used in BadNets and DBA approaches. Even if a stranger is given benign samples and images with SAB triggers to distinguish, it is not easy to discriminate which image is poisonous. The SAB-generated backdoor images are challenging to discern by the human eye. Nonetheless, our method retains strong concealment from a computer's perspective, as explicated in Sections C and D of this chapter.
    
    \subsection{Bottom-95\% and Sparse-update}
        In this paper, we deployed bottom-95\% and Sparse-update techniques to enhance the accuracy and lifespan of backdoors. Bottom-95\% means that only uploading the smaller 95\% of the gradient value when updating the gradients containing malicious content. Meanwhile, benign updates are retained in the top-5\% of gradient values. As the gradients containing backdoors and those without backdoors typically differ significantly, the location of the maximum data point in the gradients containing backdoors and those without them are often different. During model aggregation, there are different gradient values from different clients for the same data point, and the benign gradients are typically smaller than the backdoor gradients. This results in the backdoor gradients dominating during aggregation, making aggregation algorithm can not remove the backdoor, allowing it to be implanted in the model. Since the backdoor is always implanted in positions where the benign gradient values are small, even after the attack is over and the benign gradients continue to update normally, the influence of the backdoor gradient will not be eliminated in a short time because benign updates do not cause significant changes in the gradient. For normal samples, uploading sparse gradients is an effective method to prevent gradient leakage attacks and membership inference attacks, and has certain defense capabilities against typical backdoor attacks. We found that if we deploy Sparse-update, our attack method has a certain degree of ASR improvement and prolongs the survival time of the backdoor. We believe that this is profit by the fact that our trigger size is comparable to the image size, small-scale gradient modifications like 20\% cannot eliminate the effects of SAB backdoors. Our trigger overlaid on the sample creates a new complete image, which allows the model to map more features to the target class when learning the SAB backdoor sample. While this approach can result in greater redundancy in our trigger, excessive training iterations will lead to overfitting. Using sparse updates is an effective way to prevent overfitting of the trigger. By discarding 20\% of the updates, the remaining 80\% of updates are sufficient for the model to recognize the backdoor. In our experiments, 20\% was found to be the optimal proportion to discard. When the proportion of discarded parameters exceeds 20\%, the effectiveness of the backdoor is noticeably reduced. Thus, even if Sparse-update gives up some parameter updates, will make Sparse-update's defense function much less effective and our backdoor will not be significantly affected.
    \subsection{STRIP}
\begin{figure}
            \centering
            \includegraphics[scale=.8]{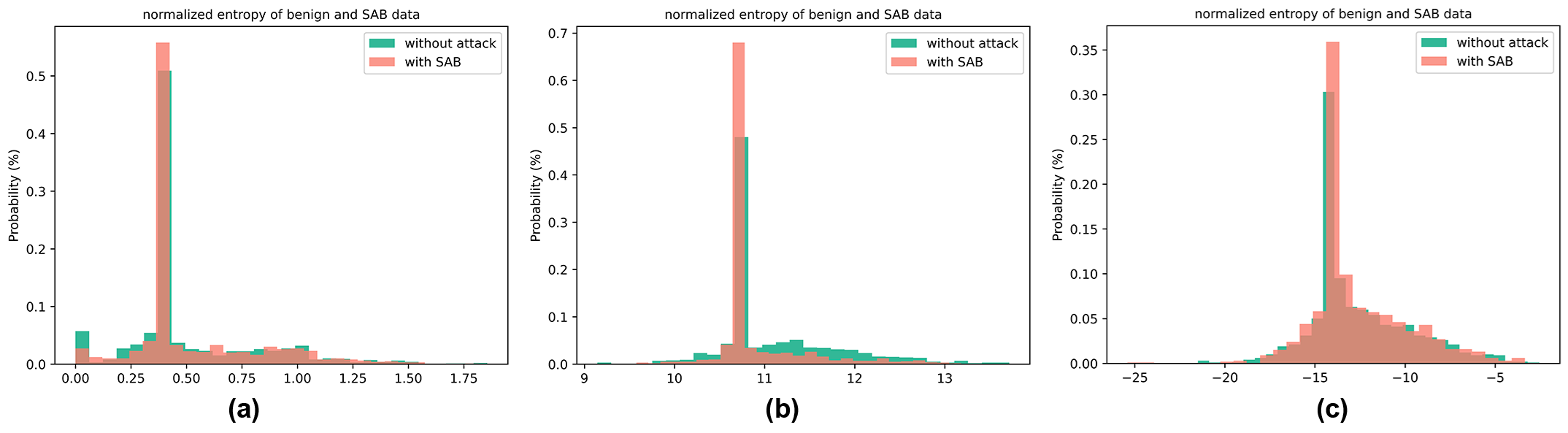}
            \caption{The distribution histograms between the poisoned dataset and the benign dataset by the SAB attack under the SAB model to the attacks on Cifar10(a), Cifar100(b), and Fashion-MNIST(c).(2-column)}
            \label{fig:14}
        \end{figure}

\begin{figure}
            \centering
            \includegraphics[scale=.8]{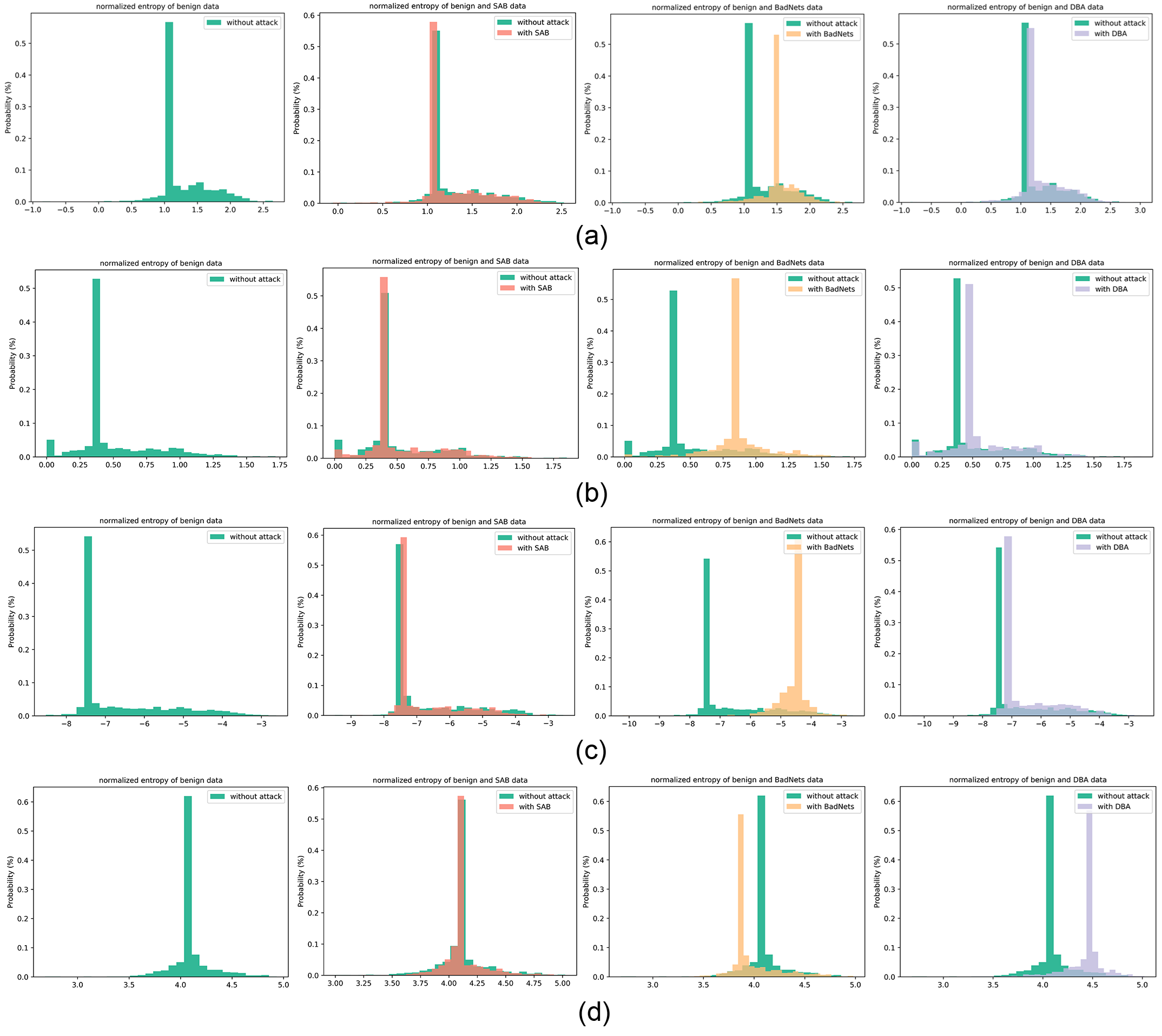}
            \caption{The distribution histograms between the poisoned dataset and the benign dataset on benign model(a), SAB model(b), BadNets(c), DBA(d) and on Cifar10.(2-column)}
            \label{fig:15}
        \end{figure}

\begin{figure}
            \centering
            \includegraphics[scale=.8]{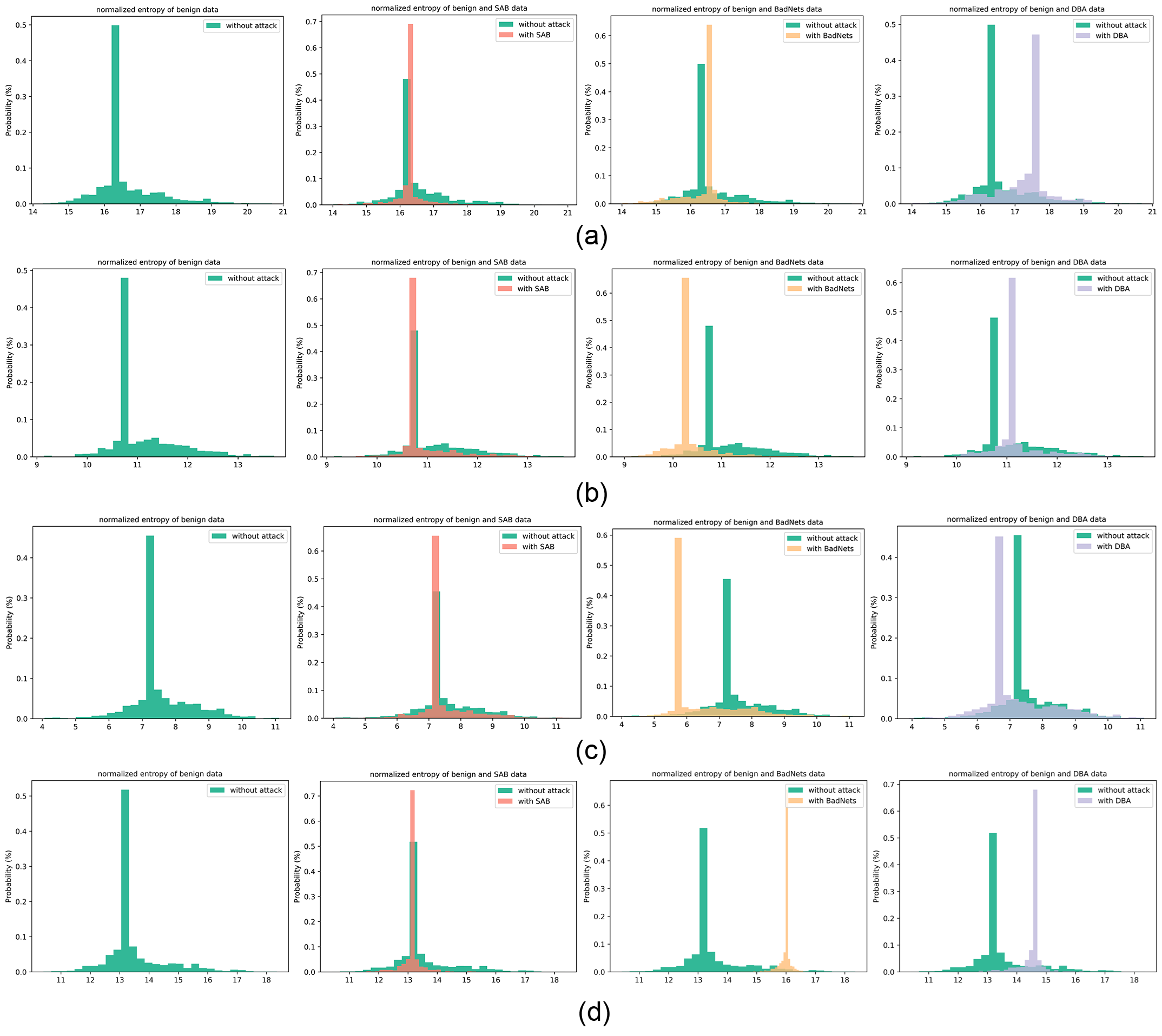}
            \caption{The distribution histograms between the poisoned dataset and the benign dataset on benign model(a), SAB model(b), BadNets(c), DBA(d) and on Cifar100.(2-column)}
            \label{fig:16}
        \end{figure}

\begin{figure}
            \centering
            \includegraphics[scale=.8]{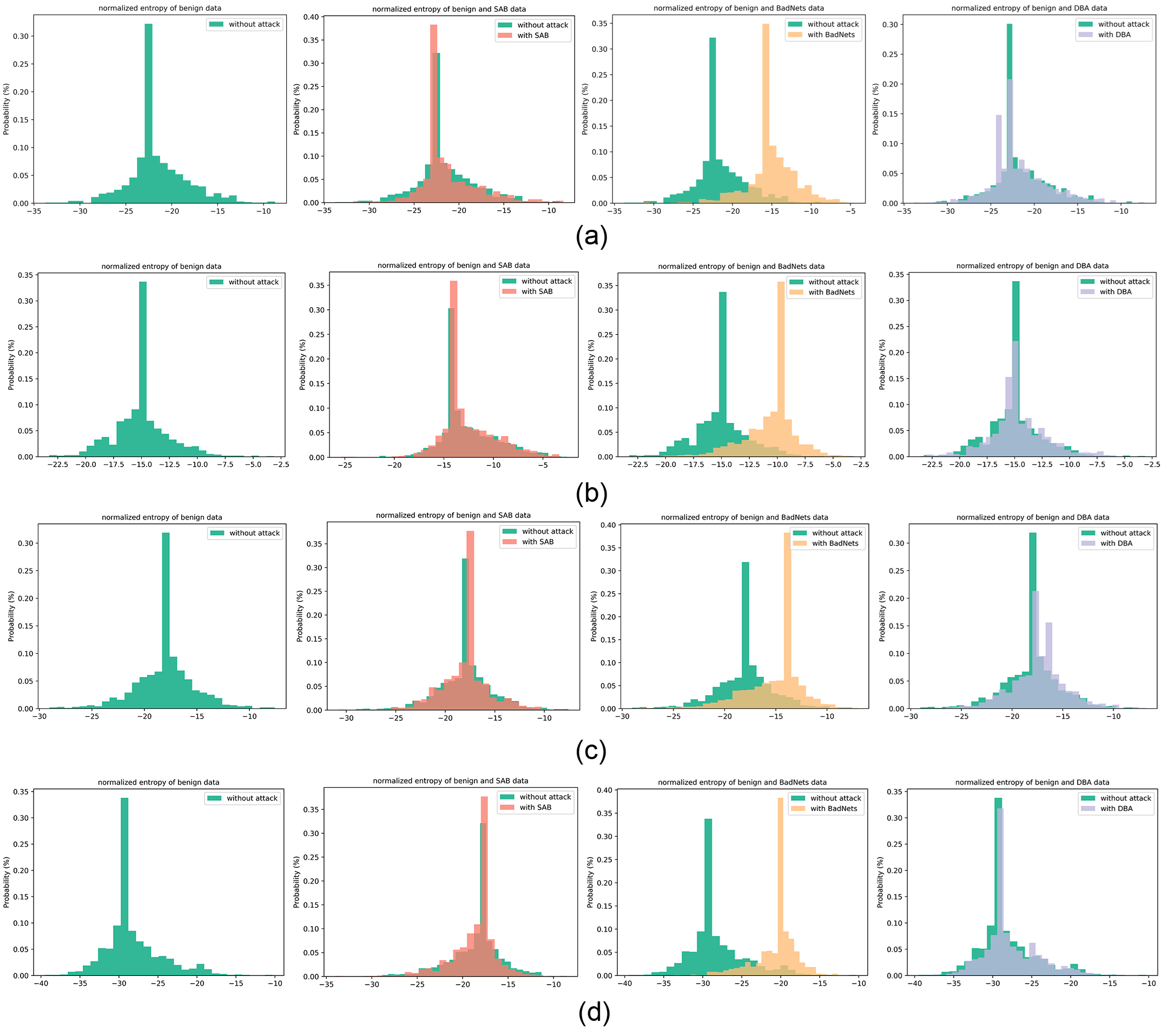}
            \caption{The distribution histograms between the poisoned dataset and the benign dataset on benign model(a), SAB model(b), BadNets(c), DBA(d) and on Fashion-MNIST.(2-column)}
            \label{fig:17}
        \end{figure}

        Comparing all model cases, after adding random strong perturbations to samples of SAB method. Model produces a set of predicted results, and the information entropy values of this set of results are used to make histograms to visualize their distribution. To prove the validity of our method against STRIP, we demonstrate that the distribution of information entropy values of our method is akin to that of benign samples. It cannot be set a threshold value explicitly to filter the images containing the SAB attack method. While BadNets and DBA methods can be filtered by setting a threshold easily. The distribution of entropy of poisoned data by SAB and clean images is similar, on the contrary the distribution of entropy of poisoned data by others’ methods and clean images is not similar. We list the comparison plots of the distribution histograms between the SAB attack dataset and the benign dataset produced under the model which is injected SAB attack in Fig. \ref{fig:14} about Cifar10, Cifar100, and Fashion-MNIST datasets. Compare SAB and the benign sample in the model with the SAB backdoor implanted. The green distribution is the benign sample distribution. Similarly, the BadNets dataset is compared with the benign dataset in the model with the BadNets backdoor to reduce the impact of the model itself on the STRIP approach. Given the varied backdoor methods in being integrated into model. We compare the distribution histograms between the poisoned dataset and the benign dataset produced by other attacks under corresponding attack model with the Cifar10, Cifar100, and Fashion-MNIST datasets to avoid the impact of the difference in the models themselves on this defense method, as Fig. \ref{fig:15}, Fig. \ref{fig:16}, and Fig. \ref{fig:17}. Accordingly, we compared each model, and the results were similar, and our SAB method always had the best results, the effectiveness of our method can be demonstrated. 
        \FloatBarrier
    \subsection{Grad-Cam}
\begin{figure}[htbp]
            \centering
            \includegraphics[scale=.8]{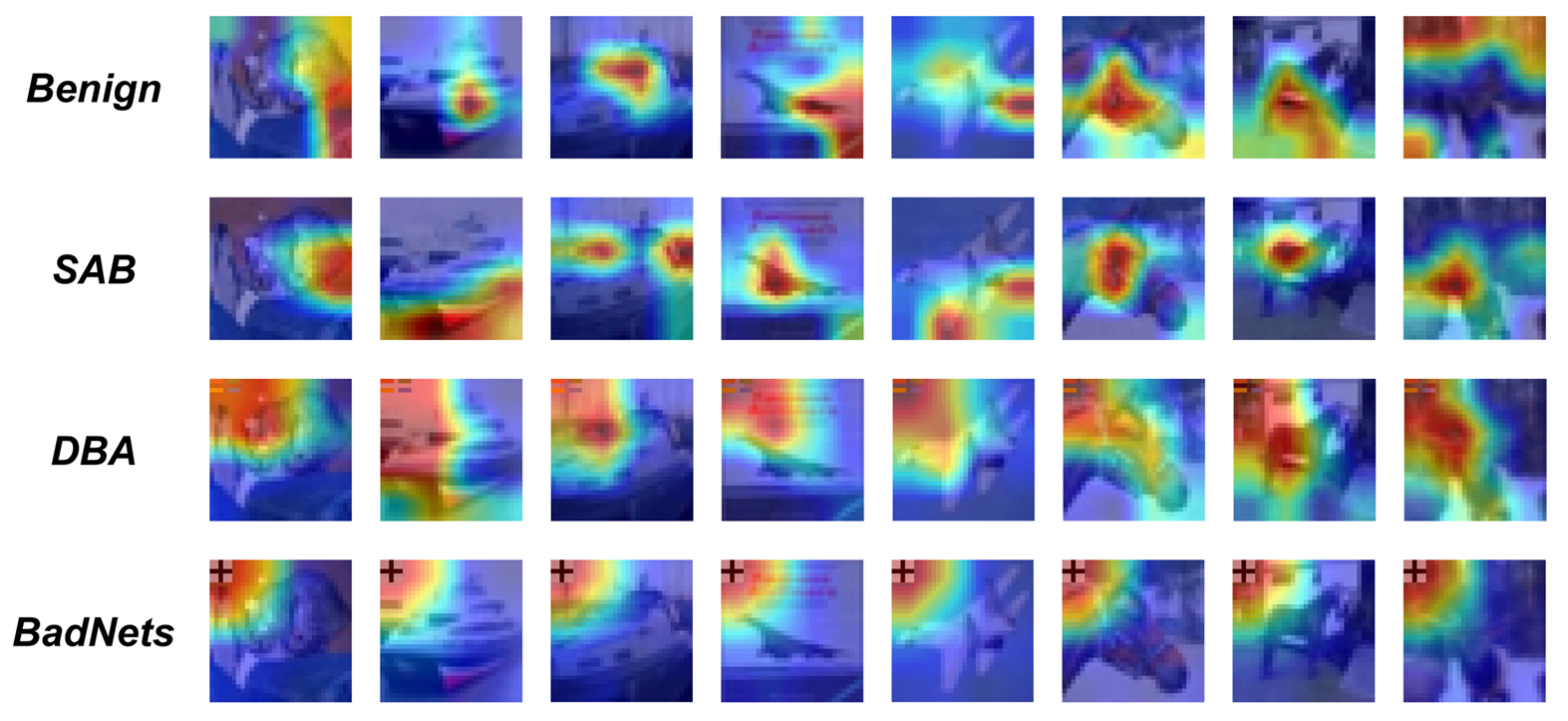}
            \caption{Grad-Cam on Benign, SAB, DBA, BadNets.(2-column)}
            \label{fig:18}
        \end{figure}

        Grad-Cam can determine the model's attention heat map by the gradient derived from the model's prediction. For some backdoor implantation methods including a fixed location, size, shape, and value of the triggers in sample, the model will focus its attention on the location of the triggers when it detects the backdoor. When the defender finds that the model focuses on a fixed area in the image when predicting, it is easy to infer that the model is embedded with a backdoor and contains a fixed trigger in the image. We show in Fig. \ref{fig:18} where model focuses its attention on the poisoned samples when SAB, BadNets, and DBA are implanted, respectively. SAB has the most decentralized attention and is therefore more complex to be determined by Grad-Cam. In contrast, the backdoors implanted by BadNets and DBA methods are focused on the trigger location by Grad-Cam and are more easily detected.
        \FloatBarrier
    \subsection{pHash}
\begin{figure}[htbp]
            \centering
            \includegraphics[scale=.8]{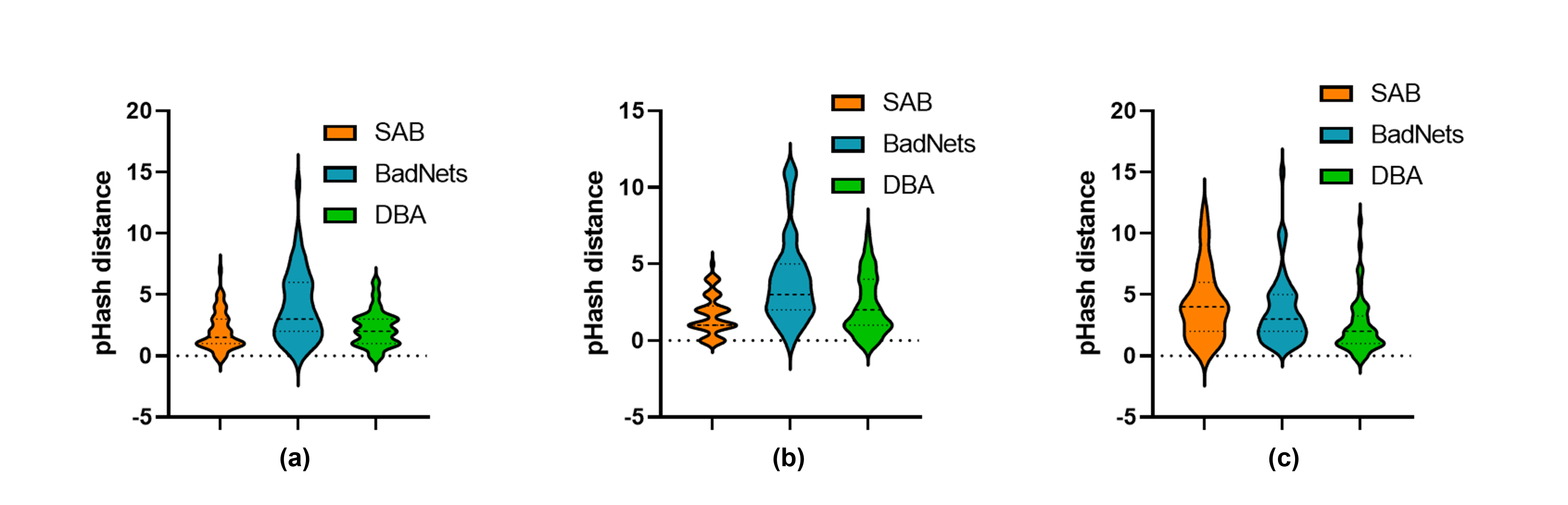}
            \caption{The distribution of pHash values on Cifar10(a), Cifar100(b), and Fashion-MNIST(c).(2-column)}
            \label{fig:19}
        \end{figure}

        The pHash represents the distance between two images, and the pHash method is closer to the human eye's inference. We choose the pHash method to calculate the distance between SAB, BadNets, DBA, and benign samples. The result exhibits the distribution of pHash values with violin plots. According to the distribution of pHash values Fig \ref{fig:19}. We find that SAB has a better effect on the 3 data sets. Nonetheless when the image has more complex pixels, the distance will be shorter. The violin plot results on the Fashion-MNIST dataset represent that the pHash value data distribution of SAB’s versus benign samples is not concentrated around the median value close to 0 as in Cifar10. We think that this is because Fashion-MNIST does not have complex pixels like CIFAR-10 or CIFAR-100, which have a solid black background and produce a larger gap, increasing the pHash value. But it is still arduous for the human eye to detect a large difference in the actual image of Fashion-MNIST.
        \FloatBarrier
\section{Conclusion}
    Today's federated learning backdoors are based on significant triggers, which can be easily identified and predicted by some defense methods. Also, these poisoned samples can be easily identified artificially. Despite recent advancements, the application of backdoor technique in the physical world remains problematic. For example, in the field of autonomous driving, when a white block appears on a sign, it can easily be considered a sign that is not conducive to the correct identification of the model. It will be replaced by a clean sign. A such significant trigger-based backdoor is strenuous to be truly deployed in the physical world. Instead, we propose an image steganography-based approach by jointly computing multiple losses to derive the most intricate trigger to detect by the eye. SAB works better and applies it in federated learning. It makes a trigger which size same as the image size and will be implanted in the bottom-95\% of the gradient of the model update, which greatly improves attack success rate of the backdoor. And it enables to be implanted faster and survive for a longer period. At the same time, thanks to the full-size triggers of SAB, the backdoor can attenuate the influence of DP on it and can evade the defenses of STRIP and Grad-Cam.

    We only encode one trigger for the image using the real label content of the image even though we can encode different triggers by changing the text content. In future work, we can try multiple triggers for one image to make more threats hidden in the model. Although we have studied the invisible triggers for application in the physical world, the triggers in this study have not experimented in the physical world. In the future, we will expand the pixels of sample to make the images clearer for real testing at the physical level.
    \FloatBarrier

\section{Acknowledgements}
We extend our sincerest appreciation to the Central Government Guides Local Science and the Technology Development Special Funds [2018]4008 for their invaluable support. Furthermore, we express our gratitude to the Science and Technology Planned Project of Guizhou Province, China under grant [2023]YB449 for their generous contributions to our research endeavors.

\bibliographystyle{cas-model2-names}

\bibliography{reference.bib}

\end{document}